\documentclass[natbib,smallextended]{svjour3}

\usepackage[table]{xcolor}
\usepackage[most]{tcolorbox}
\usepackage{multirow}
\usepackage{url}
\usepackage{graphicx}
\usepackage{pbox}
\usepackage{adjustbox}
\usepackage{array}
\usepackage{dingbat}

\newcommand*{\thead}[1]{\multicolumn{1}{c}{\bfseries #1}}

\newcolumntype{R}[2]{>{\adjustbox{angle=#1,lap=\width-(#2)}\bgroup}l<{\egroup}}
\newcommand*\rot{\multicolumn{1}{R{45}{1em}}}

\journalname{Empirical Software Engineering}

\begin{document}

\title{Testing Research Software: An In-Depth Survey of Practices, Methods, and Tools}

\author{Nasir~U.~Eisty* \and      
        Upulee~Kanewala* \and 
        Jeffrey~C.~Carver
}

\institute{
    N. U. Eisty \at
    Department of Computer Science \\
    Boise State University \\
    Boise, ID, USA \\
    \email{nasireisty@boisestate.edu}
    \and
    U. Kanewala \at
    School of Computing \\
    University of North Florida \\
    Jacksonville, FL, USA \\
    \email{upulee.kanewala@unf.edu}
    \and
    J. C. Carver \at
    Department of Computer Science \\
    University of Alabama \\
    Tuscaloosa, AL, USA \\
    \email{carver@cs.ua.edu\\}
    *Nasir Eisty and Upulee Kanewala share the first authorship.
}

\date{Received: data / Accepted: date}

\maketitle

\begin{abstract}
\textit{Context}
Research software is essential for developing advanced tools and models to solve complex research problems and drive innovation across domains.
Therefore, it is essential to ensure its correctness. 
Software testing plays a vital role in this task. 
However, testing research software is challenging due to the software's complexity and to the unique culture of the research software community.
\textit{Aims}
Building on previous research, this study provides an in-depth investigation of testing practices in research software, focusing on test case design, challenges with expected outputs, use of quality metrics, execution methods, tools, and desired tool features. Additionally, we explore whether demographic factors influence testing processes.
\textit{Method}
We survey research software developers to understand how they design test cases, handle output challenges, use metrics, execute tests, and select tools.
\textit{Results}
Research software testing varies widely.
The primary challenges are test case design, evaluating test quality, and evaluating the correctness of test outputs.
Overall, research software developers are not familiar with existing testing tools and have a need for new tools to support their specific needs.
\textit{Conclusion}
Allocating human resources to testing and providing developers with knowledge about effective testing techniques are important steps toward improving the testing process of research software. 
While many industrial testing tools exist, they are inadequate for testing research software due to its complexity, specialized algorithms, continuous updates, and need for flexible, custom testing approaches.
Access to a standard set of testing tools that address these special characteristics will increase level of testing in research software development and reduce the overhead of distributing knowledge about software testing.
 \keywords{Software Testing \and Research Software Engineering \and Software Quality \and Survey \and Interview \and Software Engineering}
\end{abstract}

\section{Introduction}
\label{sec:introduction}
\emph{Research software} is software developed and used to support research~\citep{8588655}. 
This type of software ranges from simple scripts and tools to complex systems designed to support research workflows, complex computations, data analysis, simulation, and modeling~\citep{5337646}.
This software is developed in academic or research settings in support of a wide range of domains, including but not
limited to biology, chemistry, physics, engineering, social sciences, and humanities~\citep{Kelly08thechallenge}. 
Such research software is essential as it accelerates discoveries, ensures the reproducibility and transparency of scientific work, and enables the management and analysis of complex data~\citep{10.1145/1882362.1882383}. 
It allows researchers to process large datasets, run simulations, and test hypotheses quickly, making it possible to achieve timely breakthroughs and validate findings~\citep{10078171}. Moreover, it supports collaboration across disciplines and geographies, enabling more comprehensive studies and innovative solutions~\citep{10071971}.
Research software is critical in modern science as it enables the development of cutting-edge technologies and solutions to address complex real-world challenges such as simulations that analyze the spread of COVID-19~\citep{kerr2021covasim}, software to decode raw brain data~\citep{HBM:HBM23730} and models for analyzing runoff and water quality in urban environments~\citep{epa.gov_swmm}. Thus, research software is essential in enabling scientific discoveries and addressing key societal needs. 

However, research software has several requirements that must be met.
Research software typically requires its developers and users to possess expertise and specialized knowledge~\citep{Feitosa_undated-ls}. 
Research software may also require significant resources, including computing power and storage capacity. In addition, the results produced by research software may be essential components of peer-reviewed publications~\citep{Miller1856}. 
Therefore, researchers need to trust the reliability and effectiveness of this software.
In addition, the culture of research software development is unique in its emphasis on collaboration, openness, and iterative improvement, driven by the needs of the scientific community~\citep{arvanitou2021software}, similar to an agile methodology or open-source community. 
Unlike commercial software development, where the focus may be on user experience and marketability, research software development is more concerned with creating tools that can handle complex, evolving datasets and support cutting-edge research~\citep{storer2017bridging}. 

Software testing can help ensure these qualities and is important to research software for several reasons~\citep{5999647, eisty2022}. 
First, testing plays a significant role in ensuring the accuracy of the software and the results it produces~\citep{doi:10.1177/1094342004048534}. 
Second, testing and validation help ensure other researchers can reproduce results~\citep{8565942}.
Third, testing can help identify and address faults and issues early in development, saving overall time and resources~\citep{10.1007/978-3-540-69389-5_34}.

However, there are challenges in testing research software. First, research software can be highly complex, involving multiple components, data sources, and algorithms, which makes it difficult to design effective tests~\citep{KANEWALA20141219}. 
Second, research software developers can use various programming languages, libraries, and tools, which makes it challenging to test across platforms and environments~\citep{5069155}.
Third, research software often uses large, complex datasets that are highly variable and may contain errors or inconsistencies~\citep{eisty2022}. 
Fourth, research software may not have clearly defined requirements, which makes it difficult to design tests that adequately cover all expected behaviors~\citep{HEATON2015207}. 
Finally, research software is often developed by researchers who have limited resources and software development expertise, which makes it challenging to invest sufficient time and resources in comprehensive testing and validation~\citep{article58698}.

Industrial software testing tools and techniques are often inadequate for research software because research software is typically more complex, domain-specific, and continuously evolving~\citep{eisty2022}. 
Unlike industrial software, which has clear specifications and predictable behavior, research software involves specialized algorithms and models tailored to unique scientific problems, making it difficult for standard testing tools to be effective. 
Therefore, achieving a test oracle is difficult in research software~\citep{KANEWALA20141219}. Additionally, research software is frequently updated as new scientific insights emerge, requiring testing methods that can quickly adapt to changing requirements~\citep{10.1145/2896971.2896981}.

In our previous work, we surveyed research software developers about their understanding and utilization of software testing~\citep{eisty2022}. 
Our findings show research software developers possess moderate confidence in their testing knowledge, with most reporting an average or higher understanding of the testing concepts relevant to their projects. 
We drew upon the results of this study, along with several other previous works, to formulate the research questions presented in Section~\ref{sec:researchQuestions}.

To expand on the results of the study described above~\citep{eisty2022} and from our other previous studies~\citep{10.1007/978-3-030-50436-6_33,KANEWALA20141219, 8588655}, the current survey provides a deeper understanding of the technical issues related to software testing identified in those earlier studies. 
Specifically, this current study investigates how research software developers approach test case design, the difficulties with determining expected outputs, the adoption of software and test quality metrics, test case execution methodologies, tools used, and the specific attributes desired in these tools. 
We also explore whether certain demographic characteristics are related to variations in the testing processes. 
Section~\ref{sec:researchQuestions} describes our specific research questions.

The primary contributions of this paper, beyond those from our earlier works, are:
\begin{itemize}
    \item The first study to conduct an in-depth survey about test case design, challenges with expected outputs, use of quality metrics, execution methods, tools, and desired tool features for research software.
    \item An overview of the specific characteristics of the testing process for research software, including the level of automation, the systematic/ad-hoc nature of the testing process, and the presence of test documentation.
    \item A description of the challenges encountered when testing research software, their root causes, and the relative importance as ranked by the respondents.
    \item An analysis of methods and tools used for various testing activities such as test input design, test execution, determining test output correctness, and measuring test quality.
\item Identification of demographic characteristics within research software projects that exhibit differences in their testing processes.
\end{itemize}

\section{Background}
\label{sec:background}
Research software enables advances with significant societal impact in many different fields. For instance, Covasim~\citep{kerr2021covasim} is a stochastic agent-based simulator used to analyze the spread of COVID-19. It provides projections on key indicators such as infection rates and peak hospital demand while also allowing researchers to evaluate the potential impact of interventions like social distancing, school closures, testing, and vaccination strategies. This kind of software has been essential in guiding public health decisions during the pandemic.Similarly, Braindecode~\citep{HBM:HBM23730} is an open-source Python toolbox designed for decoding raw brain data using deep learning models. It provides powerful tools for preprocessing and visualizing electrophysiological data such as EEG, ECoG, and MEG, as well as deep learning architectures that facilitate advanced neurological research. Another example is SWMM~\citep{epa.gov_swmm}, a dynamic hydrology-hydraulic simulation model used to analyze runoff and water quality in urban environments. It enables long-term simulations that help researchers understand and manage water systems effectively. These examples highlight research software's essential role in enabling scientific discoveries and addressing key societal needs.

Research software is complex because it must handle specialized, domain-specific algorithms and models, often across multiple disciplines~\citep{allan2008managing}. 
It must process large, intricate datasets, adapt to evolving scientific requirements, and deal with non-deterministic processes, making validation challenging~\citep{8588655}. 
The lack of standardization, combined with limited resources and the need for high precision further adds to its complexity~\citep{eisty2022}. 
As a result, developing and maintaining research software requires advanced technical expertise and flexibility to meet the dynamic needs of cutting-edge research. Developers often work closely with researchers to tailor software to solve specific scientific problems, prioritizing accuracy, reproducibility, and transparency. This culture also values open-source principles, with a strong emphasis on sharing code and on methodologies that foster collaboration and accelerate scientific progress. Thus, to ensure that research software is reliable and effective, it must be tested and validated through rigorous software engineering practices.

Our previous survey done on research software developers shows that these developers maintain clear testing objectives and see the utility of various testing techniques. However, they face challenges in testing research software, including test case design, resource constraints, external dependencies, and a lack of expertise. Furthermore, it is not always easy for them to incorporate Commercial/IT testing techniques~\citep{eisty2022}. This study aims to gain a comprehensive understanding of software testing within research software development, encompassing aspects such as developers' knowledge, current testing practices, testing difficulties inherent in research software, challenges in adapting existing testing methods, and potential enhancements to the testing process.  

\section{Research Questions}
\label{sec:researchQuestions}

This section describes the research questions that drive the study.
To motivate each question, we provide the relevant background from our prior work and from the literature.

\vspace{4pt}
\noindent
\textbf{RQ1: What are the characteristics of the research software testing process?}

\noindent 
Because the development environment for research software differs from that of traditional software, developers can adopt different testing processes. 
The results of our previous study~\citep{eisty2022} show research software developers most commonly use \emph{unit testing}, followed by \textit{integration} and \textit{system} testing. 
For specific testing techniques, they most commonly use assertion checking and performance testing. 
    
In addition to validating the results of our previous survey with a larger sample, we want to understand whether the testing process of research software is ad-hoc, systematic, manual, or automated and whether the test requirements are documented. 
Answers to these questions will provide an overall understanding of the testing process used in the development of research software.

\vspace{4pt}
\noindent
\textbf{RQ2: What challenges do developers face throughout the testing process of research software?}

\noindent
There are many challenges to testing research software. 
Our previous survey~\citep{eisty2022} found the most common challenges were test case design, lack of resources, and external dependencies.
Similarly, another of our previous studies identified four categories of challenges: (1) developing test cases, (2) producing expected test outputs, (3) test execution, and (4) interpreting test results~\citep{KANEWALA20141219}.

However, none of these studies provide insights into the relative frequencies of these challenges. 
Therefore, in this study, we asked the respondents to rank testing tasks according to their difficulty level. We also asked them to explain why a given task is challenging. We believe this question would allow us to gain more insight into why certain testing tasks are challenging for research software.

\vspace{4pt}    
\noindent
\textbf{RQ3: How do research software developers design test inputs?}

\noindent
Because our previous work~\citep{eisty2022, 10.1007/978-3-030-50436-6_33} found test input design to be one of the most challenging tasks in research software testing, we wanted to learn more about how developers create test inputs and what types of tools they use.

\vspace{4pt}
\noindent
\textbf{RQ4: What specific challenges do research software developers face when determining the expected output of test cases?}

\noindent
Another challenge identified in our previous work is determining the expected test outputs~\citep {KANEWALA20141219,10.1007/978-3-030-50436-6_33}. 
Therefore, to better understand this challenge, including its frequency and the reasons it occurs, we included this research question in this study.

\vspace{4pt}
\noindent
\textbf{RQ5: What metrics do research software developers use to measure software quality and test quality of tests?}

\noindent
In traditional software, metrics such as code coverage and bug density provide insight into software quality. 
However, it is not clear whether developers of research software use these metrics.
In one of our studies~\citep{8588655} we found even though software quality metrics are crucial for research software, developers did not use them adequately.
In this study, we wanted to determine whether that result held with a larger sample of developers and a few years later. 

\vspace{4pt}
\noindent
\textbf{RQ6: How do research software developers execute their tests?}

\noindent
Test execution consumes a lot of resources, especially for long-running tests. 
Tests for research software often face this challenge because of the software's complexity~\citep{eisty2022}. 
Therefore, research software developers often have to compromise the amount of testing to match the available resources.
Developers can use techniques like test case selection and test case prioritization to reduce the test execution cost. 
While these techniques are common in industry, it is unclear whether research software developers use them to reduce test execution costs~\citep{HEATON2015207}. 
Therefore, this question helps us gain a better understanding of how research software developers execute their tests with limited resources. 

\vspace{4pt}
\noindent
\textbf{RQ7: What testing tools do research software developers use and what are their limitations?}

\noindent
Software testing research has produced tools that automate software testing tasks, including test input generation, test execution, test selection/prioritization, and test metric collection. 
While industrial developers use these types of tools, it is unclear whether research software developers utilize them~\citep{HEATON2015207}. 
Therefore, we included this question to identify the relevant tools along with their limitations in the research software context. 
By identifying these common limitations, we are able to suggest improvements to existing tools.

\vspace{4pt}
\noindent
\textbf{RQ8: What features should a testing tool specifically developed for research software contain?}

\noindent
One of our previous studies~\citep {KANEWALA20141219} identified four challenges for testing research software (listed in RQ2). Most of these challenges, such as problems with producing test outputs and developing test cases, make some existing testing tools ineffective or impractical for testing research software. 
Therefore, we wanted to get the perspective of the research software developers on what features are suitable in testing tools developed specifically for testing this software.

\vspace{4pt}
\noindent
\textbf{RQ9: Are demographic characteristics of research software projects related to the testing process?}

\noindent
Research software projects differ significantly in domains, team sizes, and developer composition~\citep{8588655}. 
These factors impact the testing practices~\citep{eisty2022}. 
Answering this research question will identify whether certain domains affect the testing process more than others and whether the composition of development teams affects the testing process. 
To our knowledge, this is the first time this has been investigated.
 
\section{Methodology}
\label{sec:methodology}
The research questions described in Section~\ref{sec:researchQuestions} focus on the research software developers' perception of the testing process.
Therefore, we surveyed developers directly rather than attempting to extract this information from data mining.
This section describes the survey design, the data collection methods, and the data analysis procedure.

\subsection{Survey Design}
\label{sec:SurveyDesign}
We designed the survey around the research questions defined in Section~\ref{sec:researchQuestions}.
To make the survey feasible, we prioritized multiple-choice questions because we were concerned that too many open-ended questions would increase the time required to complete the survey and reduce the number of completed surveys.
To develop the appropriate set of multiple-choice answers for the survey questions, we interviewed 12 experts.
During these interviews, we asked open-ended versions of the survey questions.
Because the goal of these interviews was to identify appropriate multiple-choice answers for each question, we did not include the results in our final data analysis.

We then pilot-tested an initial version of the survey with five research software developers from different domains.
The goal of the pilot was to ensure the questions were clear and understandable.
Based on the feedback from the pilot, we made the following updates to the survey: (1) rephrased some questions to make them more understandable to research software developers, (2) modified multiple choice answer options to describe the possible results more accurately, and (3) rearranged and reformatted some questions to improve the readability and flow of the survey.

Figures~\ref{fig_surveyquestions} and~\ref{fig_surveyquestions2} contain the final list of questions, including demographic questions to characterize the sample and provide additional insight into the results.

\begin{figure*}
\caption{Survey questions - Part I}
\label{fig_surveyquestions}
\begin{tcolorbox}[enhanced, drop shadow]
\textbf{Demographic Questions}
    \begin{description}
        \item [Q1] Do you consider yourself to be an RSE? [yes, no]
        \item [Q2] If you have any of the following degrees, please mark the check box and indicate the field of the degree (e.g. Chemistry, Physics, Math, ...) [Bachelor, Masters, PhD, Other]
        \item [Q3] What is the domain of your scientific software project (e.g Chemistry, Physics, Math, ...)? 
        \item [Q4] How many FTEs (Full Time Equivalents, where 1 FTE is equivalent to 40 hr/week) work on your project? [$<1, 1-5, 6-20, >20$]
        \item [Q5] As part of a formal degree program, have you taken any software testing courses? [yes, no]
        \item [Q6] Mark any of the following types of training you have received related to your current software testing knowledge. [Workshops, Presentations, Reading books/articles, Tutorial, Other, None]
        \item [Q7] Have you received training specifically on the use of automated testing? [yes, no]
        \item [Q8] Are there any FTEs devoted to testing or QA in your project? [yes, no]
        \item [Q9] (If yes to Q8) In your project, how many FTEs (Full Time Equivalent) conduct testing or QA? [$<1, 1-5, 6-20, >20$]
        \item [Q10] What percentage of development time do you spend on software testing activities? [0, 1-25, 26-50, 50-75, 76-100, Unknown]
    \end{description}
\textbf{General Questions - RQ1}
\begin{description}
    \item [Q11] On a scale of 1-7 where 1 represents ad-hoc (or no process) and 7 represents a fully systematic process, indicate how systematic your testing process is. Here the testing process includes all activities from the creation of test inputs to evaluating the correctness of test outputs.
    \item [Q12] On a scale of 1-7 where 1 represents manual and 7 represents fully automated, indicate how automated your testing process is. Here the testing process includes all activities from the creation of test inputs to evaluating the correctness of test outputs.
    \item [Q13] Which of the following types of testing do you use in your project? [Unit testing, Integration testing, System testing, Regression testing, Acceptance testing, Other] (note the survey provided brief definitions)
    \item [Q14] Does your project have documented test requirements (specific elements of a software that the test cases must satisfy or cover) or test specifications (information about what scenarios to be tested)? [yes, no]
\end{description}
\textbf{General Questions - RQ2}
\begin{description}
    \item [Q15] Drag the following testing tasks into order based on their level of difficulty with 1 being the most difficult (at the top of the list). Use the text box next to each task to explain why that task is challenging, if applicable. [Test case design, Test execution, Evaluating the correctness of the test outputs, Evaluating test coverage, Evaluating test quality,
    Using testing tools, Other]
\end{description}
\textbf{Test Input Design - RQ3}
\begin{description}
    \item [Q16] Which of the following approaches do you use to design test cases? Use the space next to each approach to provide additional information, if applicable. [Manually, Automatically using tools, Ad-hoc, Other]
    \item [Q17] List any tools you use for test input generation or design.
\end{description}
\textbf{Test Outputs - RQ4}
\begin{description}
    \item [Q18] Do you face challenges when determining the expected output of test cases? [Frequently, Often, Rarely, Never]
    \item [Q19] (If the answer to Q18 is Frequently, Often, or Rarely) Explain the challenges faced when determining the expected output.
    \item [Q20] (If the answer to Q18 is Frequently, Often, or Rarely) Do challenges in determining the correct output limit the number of tests that you run? [yes, no]
    \item [Q21] (If the answer to Q20 is yes) How are the test cases limited? [Tests never executed, Tests executed less frequently, Other]
    \item [Q22] Do you face difficulties in determining suitable tolerances in the expected output? [yes, no]
    \item [Q23] (If the answer to Q22 is yes) Briefly explain the difficulties with tolerances and your current solution.
\end{description}
\end{tcolorbox}

 \end{figure*}

\begin{figure*}
\caption{Survey questions - Part II}
\label{fig_surveyquestions2}
\begin{tcolorbox}[enhanced, drop shadow]
\textbf{Test Metrics - RQ5}
\begin{description}
    \item [Q24] Which metrics do you use to measure the quality of your tests? [Statement coverage, Branch coverage, Other] (note the survey provided brief definitions)
    \item [Q25] Which of the following software quality metrics do you track, if any? [Number of defects in current software version, Number of defects in previous software versions, Problems assigned to internal development group, Problems assigned to external contractors or vendors, Turnaround time on defect corrections, None, Other]
    \item [Q26] List any tools that you use for measuring test quality.
\end{description}
\textbf{Test Execution - RQ6}
\begin{description}
    \item [Q27] When do you execute the test cases? [Nightly, With a push, At the end of the development cycle, Never, Other]
    \item [Q28] Do you use any tools for test execution? [Yes, No]
    \item [Q29] (If the answer to Q28 is yes) List any tools used for test execution.
    \item [Q30] Do you do test case selection/prioritization when resources are limited? [yes, no]
    \item [Q31] (If the answer to Q30 is yes) How do you perform test case selection/prioritization?
    \item [Q32] (If the answer to Q30 is yes) Do you use any tools for test selection/prioritization?
\end{description}
\textbf{Testing Tools \& Limitations- RQ7}
\begin{description}
    \item [Q33] List any tools used to support any aspect of the testing process that you have not mentioned previously in this survey.
    \item [Q34] What limitations do current testing tools have that make them difficult to use?
\end{description}

\textbf{Desired Features in Testing Tools - RQ8}
\begin{description}
    \item [Q35] Based on your experience with scientific software and having now completed this survey, if a testing tool was developed specifically for testing scientific software, what features should it contain?
\end{description}
\end{tcolorbox} \end{figure*}

\subsection{Data Collection}
We encoded the questions shown in Figures~\ref{fig_surveyquestions} and~\ref{fig_surveyquestions2} into the Qualtrics survey tool\footnote{https://www.qualtrics.com/}.
To obtain broad results, we targeted a wide sample of research software developers via the following distribution channels:
\begin{itemize}
    \item United States Research Software Engineer Association (US-RSE) Slack and email list
    \item Society of Research Software Engineering (UK) Slack
    \item IDEAS-ECP email list
    \item Better Scientific Software (BSSW) email list
\end{itemize}
Given that the previous survey~\citep{eisty2022} was two years old and was anonymous, we could not solicit the same participants for this survey.
However, because we used some of the same mailing lists, it is likely there are some common respondents.

\subsection{Data Analysis}
We first excluded any responses that lacked sufficient answers to be valuable.
We kept all responses that included answers to the quantitative questions.
We then analyzed the remaining 131 responses.

Most of the questions asked respondents to choose from multiple-choice answers.
For these questions, we computed frequency distributions based on the responses.

The remaining questions were free-response, resulting in qualitative data. 
There were two types of free-response questions.
The first type (Q17, Q26, Q29, and Q33) asked respondents to list various tools they used. 
To analyze these questions, we extracted unique tool names and counted them.
The second type (Q19, Q23, Q31, Q32, Q34, and Q35) produced data that required a more formal qualitative analysis. 
To analyze this data, we employed the following steps.
First, one author performed the initial coding of all responses to these questions.
Then, the other two authors each independently coded half of the data using the same approach as the initial coder, but creating their own set of codes.
At this point, we had two sets of codes for each question: those from the first coder and those from the second coder.
Then, each author reviewed the two sets of codes from $1/3$ of the questions to identify whether any discrepancies existed.
Finally, the three authors met to discuss and resolve any discrepancies identified during the previous steps.
In the end, all three authors agreed on the final coding.
We then used the results of the coding process to perform similar types of frequency analyses as done for the quantitative data. 
\section{Results}
\label{sec:results}
We organize this section around the research questions and survey questions.
Note that not all survey questions received the same number of responses.
Throughout this section, ``Q\#'' refers to the survey question number from Figures~\ref{fig_surveyquestions} and~\ref{fig_surveyquestions2}.
For each table, we include a \textit{total} row.
Many of the questions allowed each respondent to provide multiple answers.
In this case, the total at the bottom of the table and the corresponding percentages are based on the total number of answers given, not the number of respondents.
For clarity, the table captions contain the number of respondents.
In cases where the total row is larger than this number, it means the question allowed multiple answers for each respondent.

\subsection{\textbf{Demographics}}
For context, we first provide an overview of the key demographics of the respondents.
This section includes survey questions from Q1 to Q10. 

\subsubsection{Research software engineer (RSE)}
For Q1, the majority of the respondents, 82\% [108/131], self-identify as a RSE.

\subsubsection{Educational background}
Table~\ref{table:Degrees} contains the responses to Q2.
Respondents could provide multiple answers.

\begin{table}[!htb]
    \centering
    \caption{Types and domains of degrees (131 Respondents)}
    \label{table:Degrees}
    \begin{tabular}{c|c|c}
\thead{Degree (Q2)} & \thead{Count} & \thead{\%}\\
\hline
Bachelor & 81 & 32\% \\
Masters & 78 & 31\% \\
PhD & 93 & 37\% \\
\hline
\textbf{Total} & \textbf{252} & 100\%\\
\end{tabular}     \vspace{5pt}
    \begin{tabular}{c|c|c}
\thead{Domains of Degrees (Q2)} & \thead{Count} & \thead{\%} \\
\hline
Science & 102 & 40\% \\
Computing & 61 & 24\%\\
Engineering & 55 & 21\% \\
Math & 32 & 12\% \\
Humanities & 6 & 2\%\\
Social Science & 2 & 1\% \\
\hline
\textbf{Total} & \textbf{258} & \textbf{100\%}
\end{tabular} \end{table}

\subsubsection{Testing background}
Table~\ref{table:TestingBackground} shows information about the respondents' testing background, such as taking formal testing courses as part of their degrees (Q5) and any training they have received on automated testing (Q7).
Table~\ref{table:TestingTraining} provides information about the respondents' sources of training in software testing (Q6).

\begin{table}[!htb]
    \centering
    \caption{Testing background (131 responses)}
    \label{table:TestingBackground}
    \begin{tabular}{l|c|c}
\thead{Formal Testing Courses (Q5)} & \thead{Count} & \thead{\%} \\
\hline
Yes & 17 & 13\% \\
No & 114 & 87\% \\
\hline
\textbf{Total} & \textbf{131} & \textbf{100\%}
\end{tabular}     \begin{tabular}{l|c|c}
\thead{Automated Testing (Q7)} & \thead{Count} & \thead{\%} \\
\hline
Yes & 39 & 30\% \\
No & 92 & 70\% \\
\hline
\textbf{Total} & \textbf{131} & \textbf{100\%}
\end{tabular} \end{table}

\begin{table}
    \centering
    \caption{Source of training (131 responses)}
    \label{table:TestingTraining}
    \begin{tabular}{l|c|c}
\thead{Source (Q6)} & \thead{Count} & \thead{\%}\\
\hline
Reading books/articles & 119 & 44\% \\
Tutorials & 78 & 29\% \\
Workshops & 53 & 20\% \\
Other & 12 & 4\% \\
None & 9 & 3\% \\
\hline
\textbf{Total} & \textbf{271} & \textbf{100\%}
\end{tabular} \end{table}

\subsubsection{Project characteristics}
Table~\ref{table:ProjectDomain} shows the project domains from Q3.
Table~\ref{table:ProjectFTEs} shows how many FTEs the respondents' projects had (Q4).
When asked whether any of those FTEs were devoted to testing or QA (Q8), only 27\% [36/131] responded \textit{Yes}.
Table~\ref{table:TestingFTEs} shows how many testing FTEs the respondents' projects had (Q9).
Table~\ref{table:TimeSpentTesting} shows how much time the respondents spent on testing activities (Q10).

\begin{table}
    \centering
    \caption{Project domain (131 responses)}
    \label{table:ProjectDomain}
    \begin{tabular}{l|c|c}
\thead{Project Domain (Q3)} & \thead{Count} & \thead{\%}\\
\hline
Science & 85 & 50\% \\
Math & 27 & 16\% \\
Computer Science & 26 & 15\% \\
Engineering & 13 & 8\% \\
Other & 18  & 11\% \\
\hline
\textbf{Total} & \textbf{169} & \textbf{100\%}
\end{tabular} \end{table}
\begin{table}
    \centering
    \caption{Project FTEs (130 responses)}
    \label{table:ProjectFTEs}
    \begin{tabular}{c|c|c}
\thead {Project FTEs (Q4)} & \thead{Count} & \thead{\%}\\
\hline
$<$ 1 & 19 & 15\% \\
1 - 5 & 77 & 59\% \\
6-20 & 19 & 15\% \\
$>$ 20 & 15 & 12\% \\
\hline
\textbf{Total} & \textbf{130} & \textbf{100\%}
\end{tabular} \end{table}

\begin{table}[!htb]
    \centering
    \caption{Testing FTEs (36 responses)}
    \label{table:TestingFTEs}
    \begin{tabular}{c|c|c}
\thead {Testing FTEs (Q9)} & \thead{Count} & \thead{\%}\\
\hline
$<$ 1 & 11 & 31\%\\
1 - 5 & 24 & 67\% \\
6-20 & 0 & 0\% \\
$>$ 20 & 1 & 2\% \\
\hline
\textbf{Total} & \textbf{36} & \textbf{100\%}
\end{tabular} \end{table}

\begin{table}
    \centering
    \caption{Time spent testing (131 responses)}
    \label{table:TimeSpentTesting}
    \begin{tabular}{c|c|c}
\thead {Time Spent Testing (Q10)} & \thead{Count} & \thead{\%}\\
\hline
1-25\% & 76 & 58\% \\
26-50\% & 42 & 32\% \\
51-75\% & 9 & 7\% \\
76-100\% & 1 & 1\% \\
Unknown & 3 & 2\% \\
\hline
\textbf{Total} & \textbf{131} & \textbf{100\%}
\end{tabular} \end{table}

\subsection{\textbf{RQ1: Characteristics of the testing process}}
The data in this section focuses on RQ1, which includes survey questions from Q11 to Q14.

\subsubsection{Characteristics of the testing process}
\begin{figure}[!htb]
    \centering
    \includegraphics[width=0.9\textwidth]{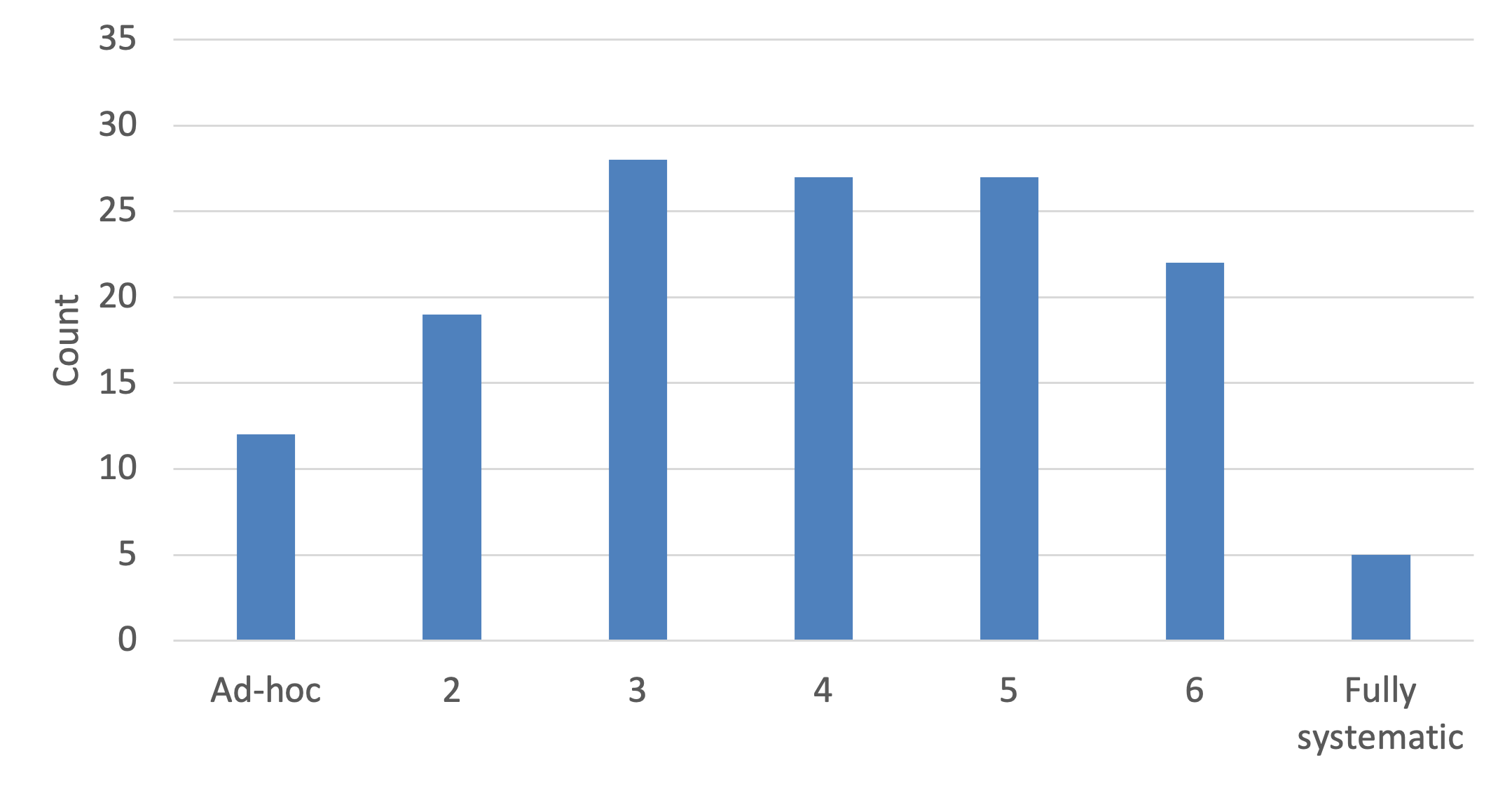}
    \caption{Whether the testing process is ad-hoc or systematic: 1 - ad-hoc and 7 - systematic (Q11)}
    \label{fig:adhoc-systematic}
\end{figure}

Regarding the systematic nature of their testing process, the responses to Q11 (Figure~\ref{fig:adhoc-systematic}) show a wide distribution, with a slight skew towards the more systematic end of the spectrum.

\begin{figure}[!htb]
    \centering
    \includegraphics[width=0.9\textwidth]{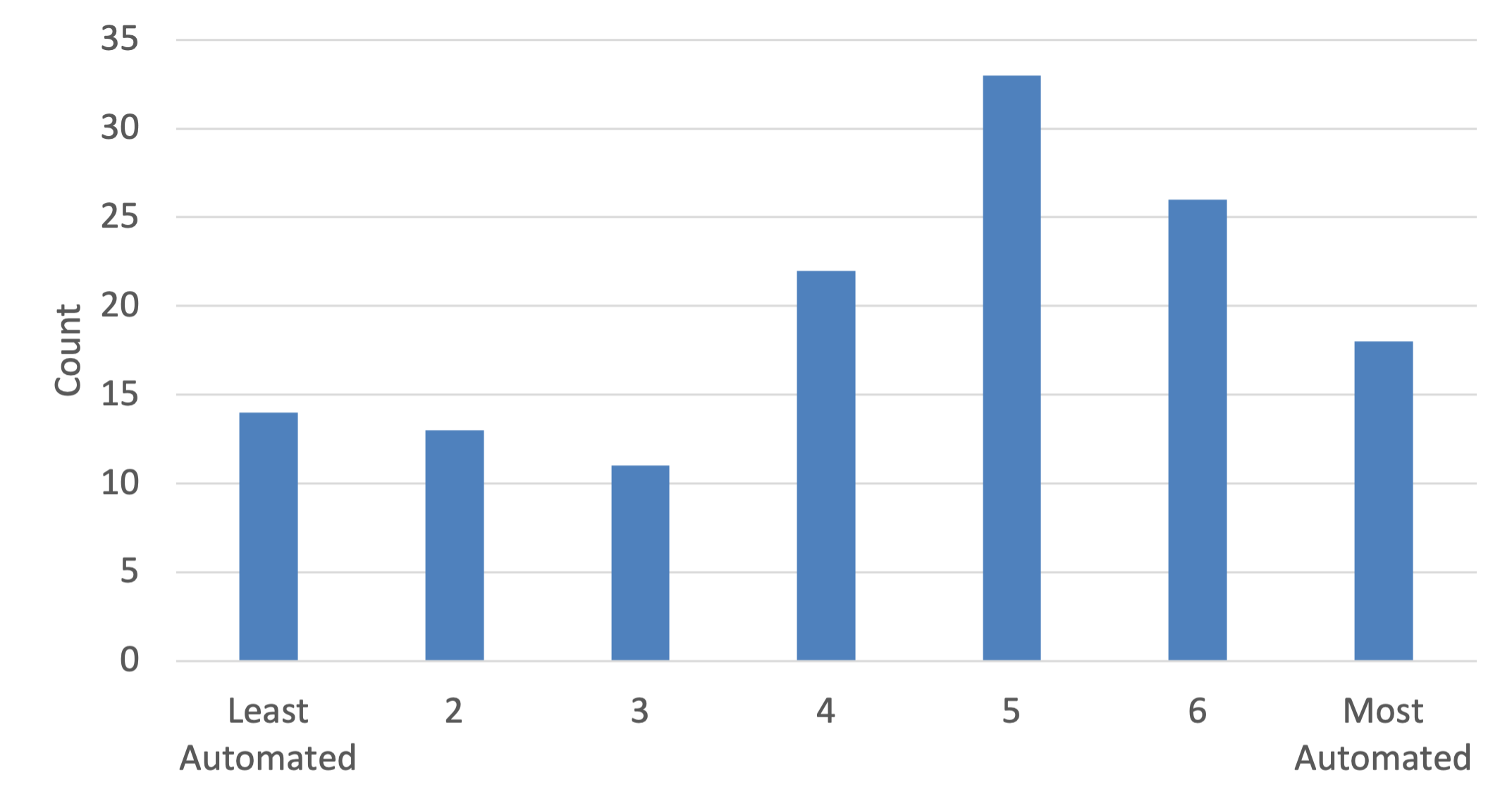}
    \caption{Whether the testing process is manual or automated: 1: least automated and 7: most automated (Q12)}
    \label{fig:manual-automated}
\end{figure}

Regarding the level of automation in their testing process, the responses to Q12 (Figure~\ref{fig:manual-automated}) again show a wide distribution, with a slight skew towards more automation.

\subsubsection{Types of testing used}
Q13 asked respondents to indicate which standard types of testing their projects used.
The survey provided definitions for each type of testing to ensure consistency.
Based on the results shown in Table~\ref{table:TestTypes} the respondents most commonly used \textit{unit testings} followed by \textit{regression testing}.

\begin{table}[!htb]
    \centering
    \caption{Types of testing used (131 responses)}
    \label{table:TestTypes}
    \begin{tabular}{l|c|c}
\thead{Testing Type/Level (Q13)} & \thead{Count} & \thead{\%} \\
\hline
Unit Testing & 116 & 26\% \\
Integration Testing & 86 & 20\% \\
System Testing & 72 & 16\% \\
Acceptance Testing & 48 & 11\% \\
Regression Testing & 106 & 24\% \\
Other & 10 & 2\% \\
\hline
\textbf{Total} & \textbf{438} & \textbf{100\%}
\end{tabular} \end{table}

\subsubsection{Documented test requirements}
In response to Q14 asking whether their projects have documented test requirements 
or test specifications, 
77 said \emph{No} and 53 said \emph{yes}.

\subsection{\textbf{RQ2: Challenges faced when testing research software}}

The data in this section focuses on RQ2, which includes survey question Q15.

\subsubsection{Testing challenges}
\begin{figure}[t]
\centering
\includegraphics[width=0.9\textwidth]{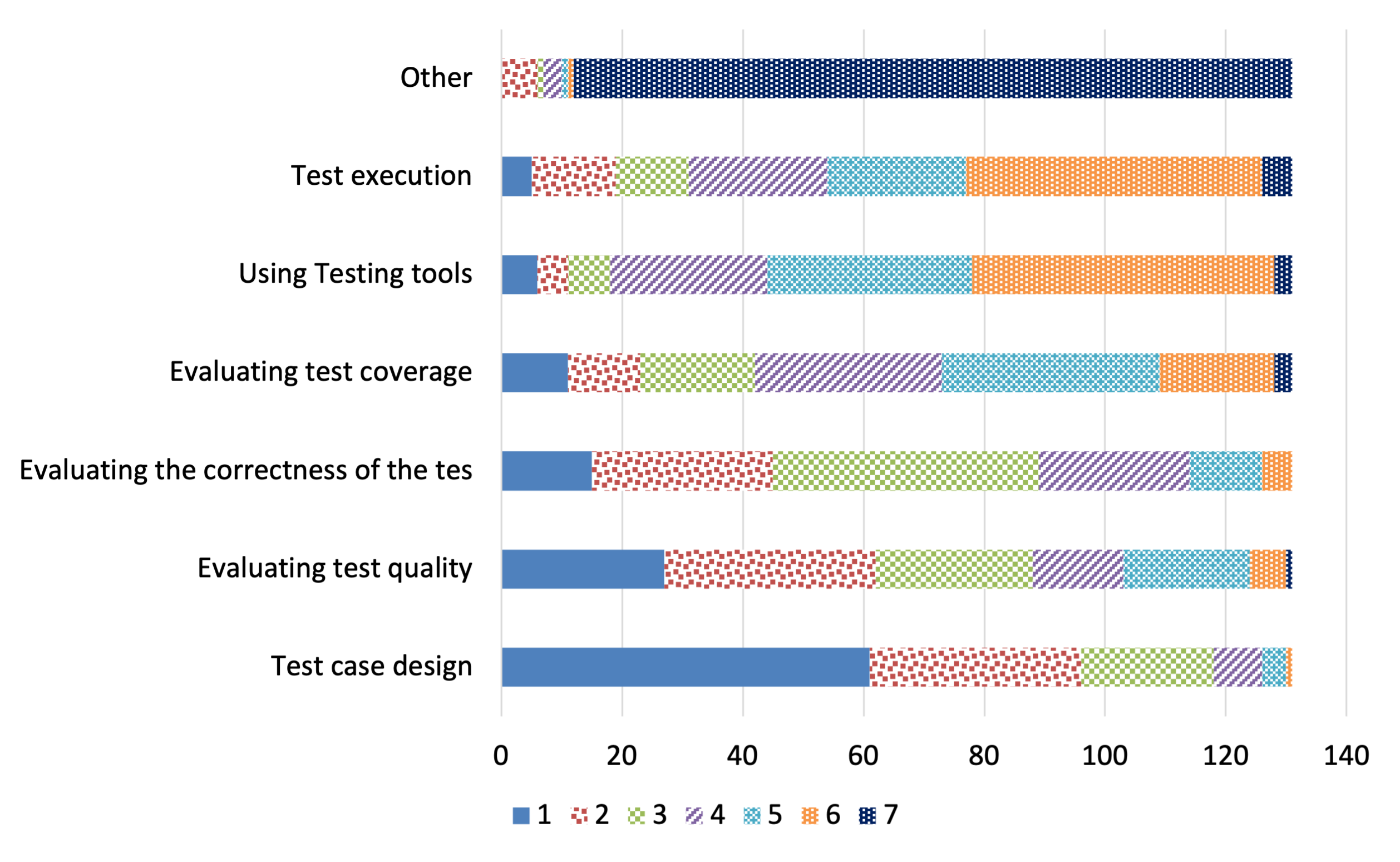}
\caption{Ranking of testing challenges. Numbers on the right indicate the ranking given to a particular testing task based on how challenging it is by the respondents. 1 - most challenging and 7 - least challenging (Q15)}
\label{fig:challengesranking}
\centering
\end{figure}

In response to Q15 (Figure~\ref{fig:challengesranking}) the respondents ranked a series of testing tasks from \textit{most difficult} to \textit{least difficult}.
They indicated \textit{Test case design}, \textit{Evaluating test quality}, and \textit{Evaluating the correctness of test outputs} were the most difficult testing tasks.

In addition to the rankings, Q15 allowed the respondents to explain their answers.
For the three most challenging activities, we provide commonly mentioned explanations, the number of respondents who gave that explanation (in parenthesis), and specific quotes from respondents.

\vspace{4pt}
\noindent\textbf{Test case design} was challenging mainly because:
\begin{itemize}
        \item it is difficult to design good test cases (10).
        \begin{itemize}
            \item ``\textit{good test design makes everything else easy}''
            \item ``\textit{Balance a quick test with physically-meaningful output, Often it's hard to know what to test}''
            \item ``\textit{Choosing what and how to test is not always obvious}''
        \end{itemize}

        \item of the \emph{oracle problem} where the expected output of a test case is unknown or difficult to define (4).
        \begin{itemize}
            \item ``\textit{real physics problems don't have exact solutions}''
        \end{itemize}
        \item of the difficulties with balancing the cost of creating test cases against the resources available for testing (4),
        \begin{itemize}
            \item ``\textit{Dealing with combinatorial explosion of inputs operations and hardware}
            \item ``\textit{balancing coverage, number of tests, resources needed for execution, and test case sensitivity is to do}''
        \end{itemize}
        \item difficult to design test cases to cover the code (6). 
        \begin{itemize}
            \item ``\textit{Achieving good coverage in a bounded amount of time is always tricky}''
            \item ``\textit{Finding all the tests needed to cover all possible statements can be hard. I do this manually}''
            \item ``\textit{More of a challenge when attempting to cover gaps in coverage}''
         \end{itemize}
    \end{itemize}

\vspace{4pt}
\noindent\textbf{Evaluating test quality} was challenging mainly because:
\begin{itemize}
        \item of the complexity of the source code (3), 
        \begin{itemize}
            \item ``\textit{Requires rigorous numerical analysis of error}''
            \item ``\textit{For scientific software, how do we know the tests reflect the proper functioning of our models, equations, etc.?, I have no idea how to do this}
        \end{itemize}
        
        \item of the lack of knowledge about testing (6), 
        \begin{itemize}
            \item ``\textit{Unclear what the metric would be}'' \item ``\textit{Insufficient understanding of quality}'' 
            \item ``\textit{Unclear we even do this}'' 
            \item ``\textit{We don't have good ways to know how good our tests actually are}
            item ``\textit{it is easy for there to be untested functions even while coverage is high}
        \end{itemize}
        \item the subjectivity of evaluation (3), 
        \begin{itemize}
            \item ``\textit{This is done ad-hoc with no real guidelines}''
            \item ``\textit{difficult due to number of devel(o)pers}
            \item ``\textit{subjectivity of quality, etc.} 
        \end{itemize}
        \item of the difficulties with determining what is actually tested (5), 
        \begin{itemize}
            \item ``\textit{You can make some automated unit tests with fake input data, but was it reasonable?}'' 
            \item ``\textit{ Questionable if usable to test detailed things like calibrations and transforms}''
            \item ``\textit{What makes a test good quality? It's not always easy to tell and something that seems to have limited value at an early stage in the development of an application may be of critical importance later.}''
        \end{itemize}
        \item of the lack of time/resources to develop quality tests (4). 
        \begin{itemize}
            \item ``\textit{Quality tests take time to develop}'' \item ``\textit{It's easy to write triv(i)al tests and takes more work to develop quality tests}''
        \end{itemize}
\end{itemize}

\vspace{4pt}
\noindent\textbf{Evaluating the correctness of outputs} was challenging mainly because:
\begin{itemize}
        \item of not knowing the expected output (6), 
         \begin{itemize}
            \item ``\textit{For integration and system tests: "correct" outputs are often not known}''
            \item ``\textit{.. For complex physical tests you can check e.g. against a reference solution that was produced e.g. with an earlier version of the code. But if the results are physically correct is difficult to evaluate.}''
            \item ``\textit{...but at the leading edge it's hard to know whether our systems are doing the right thing for complex inputs, but also hard to know whether there is emergent behavior that would not be captured by simpler test inputs. }''
        \end{itemize}
        \item of non-determinism inherent to the code (6), 
        \begin{itemize}
            \item ``\textit{Non-deterministic physics}''
            \item ``\textit{Non-deterministic parallel execution on floating point data}''
        \end{itemize}
        \item of the complexity of the underlying problem (6),
        \begin{itemize}
            \item ``\textit{can be tricky where domain knowledge is necessary}''
             \item ``\textit{ The accumulated differences on numerical calculations, makes complicated to define a test oracle}''
        \end{itemize}
        \item of the complexity of the output/s (2). 
         \begin{itemize}
         \item ``\textit{Testing floating point results is difficult across platforms and compilers let alone determining correctness bounds}''
          \item ``\textit{If a software component produces large, complex outputs, how do we know they are correct?} 
         \end{itemize}
    \end{itemize}

\subsection{\textbf{RQ3: Test input design}}
The data in this section addresses RQ3, which includes survey questions Q16 \& Q17.

\subsubsection{Design approaches}
Q16 asks respondents to indicate which test input design approaches they use.
According to the results in Table~\ref{table:TestInputDesign}, most respondents (124) used a \textit{Manual} approach. 
In explaining their responses, respondents reported the following manual approaches: \textit{designing test cases to compare with known outputs}, \textit{creating test inputs to test corner cases}, \textit{testing different application scenarios}, \textit{comparing with manually computed outputs}, and \textit{comparing with other software}. 
The respondents also manually created unit, integration, and system tests.

The second most widely used approach was \textit{ad-hoc}, specifically, \textit{creating test inputs for bug fixes}, \textit{using simple/example inputs}, and \textit{testing corner cases}. 
The third most widely used approach was \textit{automatically using tools}, specifically, \textit{using tools provided by continuous integration and continuous delivery}, \textit{unit testing tools}, and \textit{hypothesis testing tools}. 

\begin{table}[!htb]
    \centering
    \caption{Test input design techniques (131 responses)}
    \label{table:TestInputDesign}
    \begin{tabular}{l|c|c}
\thead{Test Input Design Technique (Q16)} & \thead{Count} & \thead{\%} \\
\hline
Manually & 124 & 62\% \\
Ad-hoc & 51 & 26\% \\
Automatically using tools & 22 & 11\% \\
Other & 3 & 2\% \\
\hline
\textbf{Total} & \textbf{200} & \textbf{100\%}
\end{tabular}

 \end{table}

\subsection{\textbf{RQ4: Determining the expected output}}
The data in this section addresses RQ4, which includes survey questions from Q18 to Q24.

\subsubsection{Challenges with expected outputs}
Q18 asked respondents whether they faced challenges when determining the expected output of test cases. 
Q19 provided an opportunity to explain the respone to Q18.
Tables~\ref{table:FreqTestOutputChallenges} and~\ref{table:estOutputChallenges} show the responses.

\begin{table}[!htb]
    \centering
    \caption{Frequency of challenges faced when determining the expected output of test cases (129 responses)}
    \label{table:FreqTestOutputChallenges}
    \begin{tabular}{l|c|c}
\thead{Level of Frequency  (Q18)} & \thead{Count} & \thead{\%} \\
\hline
Frequently & 27 & 21\% \\
Often & 48 & 37\% \\
Rarely & 54 & 42\% \\
Never & 0 & 0\% \\
\hline
\textbf{Total} & \textbf{129} & \textbf{100\%}
\end{tabular} \end{table}

\begin{table}[!htb]
    \centering
    \caption{Types challenges faced when determining the expected output (92 responses)}
    \label{table:estOutputChallenges}
    \begin{tabular}{l|c|c}
\thead{Challenge Category (Q19)} & \thead{Count} & \thead{\%}\\
\hline
Oracle problem & 39 & 35\% \\
Complex problem/code & 13 & 12\% \\
Non-determinism & 11 & 10\% \\
Evaluating test outputs & 10 & 9\% \\
Numerical precision & 8 & 7\% \\
Determining the source of failure & 5 & 5\% \\
Creating diverse tests & 3 & 3\% \\
Lack of domain knowledge & 3 & 3\% \\
Test creation & 3 & 3\% \\
The scale of the problem & 3 & 3\% \\
Lack of resources & 2 & 2\% \\
Diverse platforms & 2 & 2\% \\
Legacy code & 2 & 2\% \\
Other & 7 & 6\% \\
\hline
\textbf{Total} & \textbf{111} & \textbf{100\%}
\end{tabular}
 \end{table}

Most challenges are various instances of the \emph{oracle problem}. 
For example, one respondent said \emph{"We often don't have a known solution to use for the expected output. This limits the number of tests we run in that it makes the tests harder and slower to write."}. 
Another respondent said \emph{"For more complex physical tests, where no analytical solutions to the physical problem exist..."}. 
In addition, some respondents described the oracle problem as \emph{"Difficult to find reference (true) data to compare to.."} and \emph{...Rough estimates (+/- 20\%) of expected answers are possible, but precise values are not usually available."}

The next most commonly reported challenge was the \emph{complexity of the problem/code}. 
One respondent explained \emph{"One package I've written is very hard to test ([blinded]) because of the many semirings it uses, and the mask..."}. 
Another respondent said \emph{"The challenges are more due to lack of analytical solutions for complex problems and lack of understanding the physics..."}. 
A third respondent said \emph{"...(software) consists of large procedures with long argument lists."}

The third most commonly reported challenge was the \emph{non-determinism} of the software under test. 
One respondent mentioned the difficulties with \emph{"...randomization (like Monte Carlo simulations)"}.
Another respondent explained \emph{"[e]valuating ML implementations as well as operating systems is fairly non-deterministic, so determining correct is tricky."}

\subsubsection{Limiting of test cases}
In response to Q20 concerning whether the challenges in determining the correct output limit the number of tests that can be run, 70 respondents said \textit{No} and 56 said \textit{Yes}. 
For those who responded \textit{yes}, Table~\ref{table:TestCaseLimits} lists the response provided (Q21). 
The other limitations included lack of automation, more time taken to create and evaluate tests, not creating tests at all, and writing fewer tests.

\begin{table}[!htb]
    \centering
    \caption{How test cases are limited (49 responses)}
    \label{table:TestCaseLimits}
    \begin{tabular}{l|c|c}
\thead {How Test Cases Limited (Q21)} & \thead{Count} & \thead{\%} \\
\hline
Tests executed less frequently & 23 & 42\% \\
Tests never executed & 13 & 24\% \\
Other & 19 & 35\% \\
\hline
\textbf{Total} & \textbf{55} & \textbf{100\%}
\end{tabular} \end{table}

\subsubsection{Determining tolerances}
In response to Q22 concerning whether they faced difficulties in determining suitable tolerances in the expected output, 79 replied \textit{yes} and 47 replied \textit{No}.
For those that replied \textit{yes}, Q23 asked them to explain their difficulties and current solutions.
Table~\ref{table:determiningTolerance} lists the commonly reported difficulties and solutions.

\begin{table}[!htb]
    \centering
    \caption{Frequency of difficulties and solutions when determining suitable tolerances (61 responses (Q23))}
    \label{table:determiningTolerance}
    \resizebox{\linewidth}{!}{\begin{tabular}{l|c|c||l|c|c}
\thead{Difficulties [Q23]} & \thead{Count} & \thead{\%} & \thead{Solutions [Q23]} & \thead{Count} & \thead{\%}\\
\hline
Other & 19 & 31\% & Other & 8 & 29\%\\
Choosing appropriate tolerances & 11 & 18\% & Adjust tolerance & 7 & 25\%\\
External Dependencies & 10 & 16\% & Ad hoc & 6 & 21\%\\
Oracle Problem & 8 & 13\%& Modify tests & 4 & 14\%\\
Floating Point & 6 & 10\% & Use reference solution & 3 & 11\%\\
Lack of Knowledge &	5 & 8\%& &\\
System Problems & 3 & 5\% & & \\
\hline
\textbf{Total} & \textbf{62} & \textbf{100\%} & & \textbf{28} & \textbf{100\%}

\end{tabular}
 }
\end{table}

\subsection{\textbf{RQ5: Metrics for software quality and test quality}}
This section includes data to address RQ5, which includes survey questions Q24, Q25, \& Q26.

\subsubsection{Test quality}
Q24 asked respondents which metrics (from a provided list) they used to measure the quality of tests.
Table~\ref{table:TestQualityMetrics} shows the number of respondents used each metric.
Note that respondents could provide more than one answer.

\begin{table}[!htb]
    \centering
    \caption{Test quality metrics used (111 responses)}
    \label{table:TestQualityMetrics}
    \begin{tabular}{l|c|c}
\thead{Metrics (Q24)} & \thead{Responses} & \thead{\%} \\
\hline
Statement coverage & 74 & 49\% \\
Branch coverage & 28 & 19\% \\
Other & 29 & 19\% \\
None & 20 & 13\% \\
\hline
\textbf{Total} & \textbf{151} & \textbf{100\%}
\end{tabular} \end{table}

\subsubsection{Overall quality}
Q25 asked respondents which software quality metrics (from a provided list) they tracked.
Table~\ref{table:OverallQualityMetrics} shows the number of respondents that used each metric.
Note that respondents could provide more than one answer.

\begin{table}[!htb]
    \centering
    \caption{Overall quality metrics used (121 responses)}
    \label{table:OverallQualityMetrics}
    \begin{tabular}{l|c|c}
\thead{Metrics (Q25)} & \thead{Responses} & \thead{\%} \\
\hline
Number of defects in current software version & 34 & 22\% \\
Number of defects in previous software versions & 14 & 9\% \\
Problems assigned to internal development groups & 16 & 10\% \\
Problems assigned to external contractors or vendors & 0 & 0\% \\
Turnaround time on defect corrections & 8 & 5\% \\
Other & 8 & 5\% \\
None & 74 & 48\% \\
\hline
\textbf{Total} & \textbf{154} & \textbf{100\%}
\end{tabular} \end{table}

\subsection{\textbf{RQ6: Test execution}}
This section includes data to address RQ6, which includes survey questions from Q27 to Q32. 

\subsubsection{Timing of test case execution}
Q27 asked respondents when they executed test cases.
Table~\ref{table:TestExecution} shows the number of responses for each answer choice.
If the respondents chose \textit{Other}, they could explain their answer (Table~\ref{table:TestExecutionOther}).
Out of these 24 \textit{Other} responses, 14 are \textit{Ad Hoc/Manually}, 4 are \textit{On merge}, 2 are \textit{Weekly}, 1 is \textit{Always}, and 1 is \textit{Per build}.

\begin{table}[!htb]
    \centering
    \caption{Test case execution (125 responses)}
    \label{table:TestExecution}
    \begin{tabular}{l|c|c}
\thead{Time (Q27)} & \thead{Responses} & \thead{\%} \\
\hline
With a push & 78 & 46\% \\
At the end of development cycle & 43 & 25\% \\
Nightly & 27 & 16\% \\
Other & 22 & 13\% \\
Never & 1 & 1\% \\
\hline
\textbf{Total} & \textbf{171} & \textbf{100\%}
\end{tabular} \end{table}
\begin{table}[!htb]
    \centering
    \caption{Test case execution - Other responses (22 responses)}
    \label{table:TestExecutionOther}
    \begin{tabular}{l|c|c}
\thead{Reason (Q27)} & \thead{Responses} & \thead{\%} \\
\hline
Ad hoc/Manually & 14 & 64\% \\
On merge & 4 & 18\% \\
Weekly & 2 & 9\% \\
Always & 1 & 5\% \\
Per build & 1 & 5\%\\
\hline
\textbf{Total} & \textbf{22} & \textbf{100\%}
\end{tabular} \end{table}

\subsubsection{Test case selection and prioritization}
In response to Q30 regarding whether the respondents engaged in test case selection or prioritization when faced with limited resources, 56 responded \textit{yes} and 66 responded \textit{no}.
For those that responded \textit{yes}, Q31 asked them to describe their approach.
Table~\ref{table:TestCaseSelection} shows the grouped responses.

\begin{table}[!htb]
    \centering
    \caption{Test case selection/prioritization (33 responses)}
    \label{table:TestCaseSelection}
    \begin{tabular}{l|c|c}
\thead{Category (Q31)} & \thead{Responses} & \thead{\%}\\
\hline
Manually & 56 & 56\% \\
Execution time & 9 & 9\% \\
Features & 7 & 7\%\\
Ad-hoc & 5 & 5\%\\
Others & 4 & 4\%\\
Automatically & 4 & 4\%\\
System dependent & 3 & 3\%\\
Grouping tests & 3 & 3\%\\
Problem dependent & 3 & 3\%\\
Coverage & 2 & 2\%\\
Time dependent & 2 & 2\%\\
Changed files & 2 & 2\%\\
\hline
\textbf{Total} & \textbf{100} & \textbf{100\%}
\end{tabular} \end{table}

\subsection{\textbf{RQ7: Testing tools used \& limitations}}
This section includes data to address RQ7, which includes survey questions from Q33 to Q34.
In addition to RQ7, for clarity, this section includes information from all tool-related questions across all RQs (i.e., Q17, Q26, Q29, Q32, and Q33).

\subsubsection{Input generation tools}
Q17 asked respondents to list any tools they use for test input generation or design. 
Table~\ref{table:InputGeneration} lists the responses.

\begin{table}[!htb]
    \centering
    \caption{Input generation tools (35 responses)}
    \label{table:InputGeneration}
    \begin{tabular}{l|c|c}
\thead{Tool (Q17)} & \thead{Count} & \thead{\%} \\
\hline
Custom Developed & 13 & 30\% \\
Pytest & 6 & 14\% \\
Python Hypothesis & 4 & 9\% \\
gtest & 2 & 5\% \\
ad hoc & 2 & 5\% \\
None & 4 & 9\% \\
Other & 12 & 28\% \\
\hline
\textbf{Total} & \textbf{43} & \textbf{100\%}
\end{tabular} \end{table}

\subsubsection{Test tools}
Q26 asked respondents to list tools used for measuring test quality. 
Table~\ref{table:TestTools} shows the results. 
Note that respondents could provide more than one answer.

\begin{table}[!htb]
    \centering
    \caption{Test tools used (35 responses)}
    \label{table:TestTools}
    \begin{tabular}{l|c|c}
\thead{Tool (Q26)} & \thead{Responses} & \thead{\%}\\
\hline
codecov & 11 & 19\% \\
pytest & 5 & 9\% \\
gcov & 4 & 7\% \\
Coverage.py & 3 & 5\% \\
Github actions & 3 & 5\% \\
pytest-cov & 3 & 5\%\\
cobertura & 2 & 4\% \\
Coveralls & 2 & 4\%\\
SonarCloud & 2 & 4\% \\
None & 2 & 4\% \\
Other - individual answers & 20 & 35\% \\
\hline
\textbf{Total} & \textbf{57} & \textbf{100\%}
\end{tabular} \end{table}

\subsubsection{Test execution tools}
Q28 asked whether respondents used any test execution tools. Of those that responded, 102 said \textit{yes}, and 22 said \textit{no}.
For those who responded \textit{yes}, Table~\ref{table:TestExecutionTools} lists the most commonly mentioned tools provided in response to a follow-up question (Q29).

\begin{table}[!htb]
    \centering
    \caption{Test execution tools (78 responses)}
    \label{table:TestExecutionTools}
    \begin{tabular}{l|c|c}
\thead {Tool (Q29)} & \thead{Responses} & \thead{\%} \\
\hline
GitHub Actions & 24 & 30\% \\
Pytest & 22 & 27\% \\
GitLab CI & 14 & 17\% \\
Jenkins & 11 & 14\% \\
CTest & 10 & 12\% \\
\hline
\textbf{Total} & \textbf{81} & \textbf{100\%}
\end{tabular} \end{table}

\subsubsection{Test case selection/prioritization}
Q32 asked respondents to list tools for test case selection/prioritization.
Table~\ref{table:TestCasePrioritization} lists the responses.

\begin{table}[!htb]
    \centering
    \caption{Test case selection/prioritization tools (31 responses)}
    \label{table:TestCasePrioritization}.
    \begin{tabular}{l|c|c}
\thead {Tools (Q32)} & \thead{Count} & \thead{\%} \\
\hline
No tools & 40 & 83\% \\
Pytest & 4 & 8\% \\
CTest & 2 & 4\% \\
CAE Fidesys & 1 & 2\% \\
UnitTest2 & 1 & 2\% \\
\hline
\textbf{Total} & \textbf{48} & \textbf{100\%}
\end{tabular} \end{table}

\subsubsection{Additional tools}
Q33 asked the respondents to list any other testing-related tools not previously mentioned in the survey. 
Table~\ref{table:AdditionalTools} lists tools mentioned by at least two respondents. 
The following tools were mentioned by only one respondent: ARM Forge (ddt) debugger, CAE Fidesys, Catch2, GitLab Runner, Mkdocs, Shell scripts, Sphinx, Tarpaulin, Bash, Buildbot, GitLab, R packages, Hypothesis, Make, Valgrind, Jenkins, Google Benchmark, Flake8, File comparison utility, Intel code coverage, TotalView, GitLab-CI, Codecov, PyCharm, Compiler static analysis, Ubuntu, FUNIT, Jacamar, CircleCI, Flawfinder, Autotools, and GDB.

\begin{table}[!htb]
    \centering
    \caption{Additional testing tools (33 responses)}
    \label{table:AdditionalTools}
    \begin{tabular}{l|c|c}
\thead{Tool (Q33)} & \thead{Responses} & \thead{\%} \\
\hline
CI & 4 & 12\% \\
pFUnit & 3 & 9\% \\
CDash & 3 & 9\%  \\
Docker & 3 & 9\%  \\
Python & 3 & 9\%  \\
Gcov & 3 & 9\%  \\
CTest & 2 & 6\%  \\
GitHub Actions & 2 & 6\%  \\
Google test & 2 & 6\%  \\
Custom made & 2 & 6\%  \\
Pytest & 2 & 6\%  \\
Other & 5 & 15\%  \\
\hline
\textbf{Total} & \textbf{34} & \textbf{100\%}
\end{tabular} \end{table}

Q34 asked respondents to describe the limitations of current testing tools.
Table~\ref{table:ToolLimitation} shows the grouped responses.

\begin{table}[!htb]
    \centering
    \caption{Limitations of current testing tools (43 responses)}
    \label{table:ToolLimitation}
    \begin{tabular}{l|c|c}
\thead{Limitations (Q34)} & \thead{Count} & \thead{\%}\\
\hline
Missing important features & 11 & 24\% \\
Others & 10 & 22\% \\
Lack of knowledge & 6 & 13\% \\
Resource limitation & 5 & 11\% \\
Usability & 4 & 9\% \\
Lack of documentation & 2 & 4\% \\
Lack of coverage & 2 & 4\% \\
Executing environments & 2 & 4\% \\
Software stack & 2 & 4\% \\
Automation & 2 & 4\% \\
\hline
\textbf{Total} & \textbf{46} & \textbf{100\%}
\end{tabular}

 \end{table}

\subsection{\textbf{RQ8: Desired features of testing tools}}
This section includes data to address RQ8, which includes survey question Q35.

Q35 allowed respondents to describe the features they would like to see in a testing tool for research software.
Table~\ref{table:DesiredToolFeatures} shows the grouped responses.
\begin{table}[!htb]
    \centering
    \caption{Desired features of testing tools (57 responses)}
    \label{table:DesiredToolFeatures}
    \begin{tabular}{l|c|c}
\thead{Features (Q35)} & \thead{Count} & \thead{\%} \\
\hline
Ease of use & 15 & 14\% \\
Others & 15 & 14\% \\
Software support & 11 & 10\%  \\
Handling round-off differences & 10 & 9\% \\
Automation & 8 & 8\% \\
Reporting and explainability & 8 & 8\% \\
Customizable & 6 & 6\% \\
Great documentation & 6 & 6\% \\
Support for tolerance & 5 & 5\%\\
History of tracking & 5 & 5\% \\
Provides examples & 5 & 5\%\\
Hypothesis/Property-based testing & 4 & 4\% \\
Hardware support & 3 & 3\% \\
Output formating & 3 & 3\% \\
Abstraction & 2 & 3\% \\
\hline
\textbf{Total} & \textbf{106} & \textbf{100\%}
\end{tabular}

 \end{table}

\subsection{\textbf{RQ9: Influence of Demographics on Results}}
This question includes data to address RQ9.
Using questions from the Demographics section of the survey (Q1-Q10), we analyze whether the way respondents described themselves had any relationship with their responses to the questions on the remainder of the survey.
For this analysis, we focused on four demographics:
\begin{itemize}
    \item Q3 - Project Domain
    \item Q4 - \# FTEs on the project (a measure of project size)
    \item Q8 - Whether there are any FTEs specifically devoted to testing
    \item Q10 - \% of development time spent on testing
\end{itemize}
We excluded Q1, Q5, and Q7 because the distribution of responses was too unbalanced to provide useful insights.
We excluded Q2 because many respondents had multiple degrees.
We excluded Q6 and Q9 because we received too few responses to these optional questions.

Using these demographics, we analyzed the remainder of the survey questions (described in the previous sections) 
During this analysis, we excluded questions that resulted in lists (Q17, Q26, Q29, and Q33) and free-response questions (Q19, Q23, Q31, Q32, Q34, and Q35).
In the subsections that follow, for each research question, we only report cases where the demographics impacted the distribution of the results. 
See Table~\ref{tab_question_mapping} in Appendix~\ref{appendix_a} for the mapping.
Since each subgroup size is different, we use percentages when discussing the results instead of absolute numbers.

\subsubsection{Q3 - Project domain}

The cross-cutting analysis in this section groups respondents based on their project domain.
In the table that follow, we use the following abbreviations: \textit{Computer Science (CS), Engineering (ENG), Math (MA), Science (SC), and Other (O).}

\begin{figure}[t]
\centering
\includegraphics[width=0.9\textwidth]{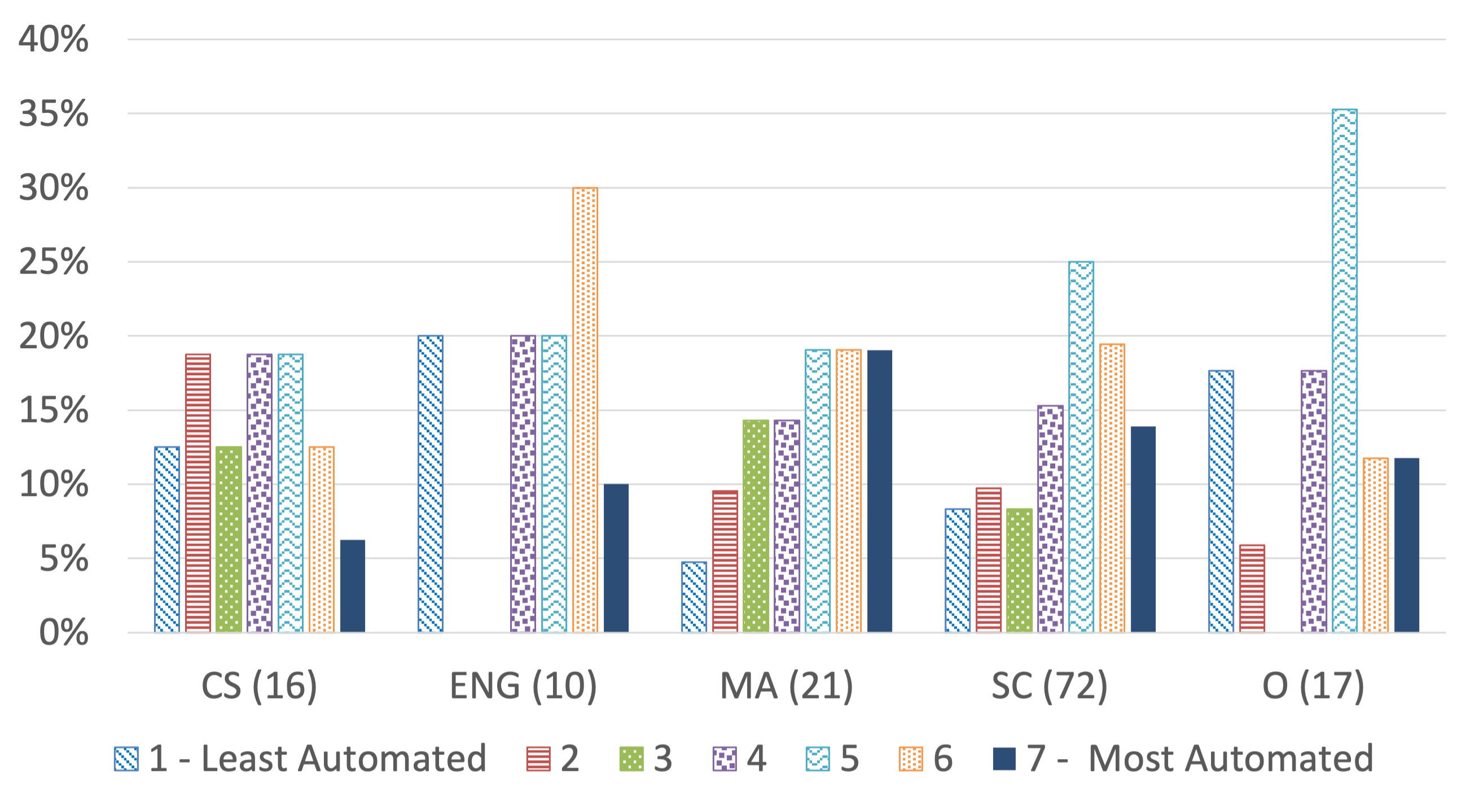}
\caption{Testing automation (Q12) vs. project domain (Q3)}
\label{fig:AutomatedVsDomain}
\centering
\end{figure}

\vspace{4pt}
\noindent
\textbf{RQ1 - Testing characteristics}
Compared to projects from other domains, \textbf{computer science} projects showed some differences in testing practices. Specifically, they were more skewed towards manual testing -- 62.5\% of the respondents from CS projects compared with 40\% - 43\% for respondents from other domains (see Figure~\ref{fig:AutomatedVsDomain}). 

\vspace{4pt}
\noindent
\textbf{RQ4 - Test outputs}

\noindent
When determining the expected output of test cases (Table~\ref{tab:DomainChallengeFrequency}), respondents from \textbf{computer science} projects were much less likely to face challenges, with 69\% \textit{rarely} facing challenges compared with 73\% of respondents from \textbf{engineering} projects and 43\% of respondents from \textbf{science} projects \textit{often} facing challenges.
Regarding whether challenges in determining the correct output limit the number of test runs (Table~\ref{tab:DomainLimitTests}), more than 50\% of the respondents from \textbf{science} projects responded \emph{Yes} while the majority of respondents from all other types of projects responded \emph{No}. 

\begin{table}[!htb]
    \centering
    \caption{Frequency of challenges with determining expected outputs (Q18) vs. project domain (Q3)}
    \label{tab:DomainChallengeFrequency}
    \begin{tabular}{l|c|c|c|c|c||c}
\thead{} & \thead{CS} & \thead{ENG} & \thead{MA} & \thead{SCI} & \thead{O} & \thead{Total} \\
\hline
Frequently & 2 (12\%) &2 (18\%)  &  8 (36\%) & 15 (20\%) & 4 (23\%) &31(22\%) \\
\hline
Often & 3 (19\%) &\cellcolor{gray!50}8 (73\%)  & 7 (32\%) & \cellcolor{gray!50}32 (43\%) & 2 (12\%)   & 52 (37\%) \\
\hline
Rarely & \cellcolor{gray!50}11 (69\%) & 1 (9\%) & 7 (32\%) &28 (37\%)  & 11 (65\%) &58 (41\%) \\
\hline
Never & 0 9(0\%) & 0 (0\%) & 0 (0\%) &0 (0\%)  & 0 (0\%) & 0 (0\%) \\
\hline\hline
Total & 16 &11  & 22  &75  &17  & 141 \\
\end{tabular} \end{table}

\begin{table}[!htb]
    \centering
    \caption{Whether challenges in determining output limit \# tests (Q20) vs. project domain (Q3)}
    \label{tab:DomainLimitTests}
    \begin{tabular}{p{.15\textwidth}|c|c|c|c|c||c}
\thead{} & \thead{CS} & \thead{ENG} & \thead{MA} & \thead{SCI} & \thead{O} & \thead{Total} \\
\hline
Yes & 4 (25\%) &5 (45\%)  & 10 (48\%) &\cellcolor{gray!50}38 (52\%)  &6 (35\%)&63 (46\%) \\
\hline
No & 12 (75\%)  &6 (52\%)  & 11 (52\%)  & 35 (48\%) &11 (35\%) & 75 (54\%) \\
\hline\hline
Total & 16 &11  &21  &73  & 17 &138  \\
\end{tabular} \end{table}

Regarding how tests are limited due to challenges in determining the correct output (Table~\ref{tab:DomainLimitTestsHow}), in \textbf{computer science} and \textbf{science} projects, the most common approach is to \emph{execute those tests less frequently}, where 50\% of the \textbf{computer science} respondents and 44\% of the \textbf{science} respondents selected this option. 
For \textbf{engineering} projects, the most common approaches were to \emph{execute those tests less frequently} (40\%) and to use \emph{other approaches} (40\%). 
For \textbf{math} projects, the most common response is to use \emph{other} approaches (45\%).

\begin{table}[!htb]
    \centering
    \caption{How challenges limit \# tests (Q21) vs. project domain (Q3)}
    \label{tab:DomainLimitTestsHow}
    \begin{tabular}{p{2cm}|c|c|c|c|c||c}
\thead{How Limited}  & \thead{CS} & \thead{ENG} & \thead{MA} & \thead{SCI} & \thead{O} & \thead{Total} \\\hline
Never Executed &1 (25\%)  & 1 (20\%) & 4 (36.3\%) &9 (23\%)  &0 (0\%)  & 15 (24\%) \\
\hline
Executed Less Frequently &2 \cellcolor{gray!50}(50\%) & \cellcolor{gray!50}2 (40\%) & 2 (18.2\%)  &\cellcolor{gray!50}17 (44\%)  &3 (0\%)  &26 (41\%)  \\
\hline
Other &1 (25\%)  &\cellcolor{gray!50} 2 (40\%) &\cellcolor{gray!50} 5 (45.5\%) & 13 (33\%)  & 1 (25\%)  & 22 (35\%) \\
\hline\hline
Total & 4 & 5  &11  &39  &4  &63  \\
\end{tabular} \end{table}

Regarding determining suitable tolerances in expected outputs (Table~\ref{tab:DomainTolerance}), the majority of respondents from \textbf{engineering} (73\%), \textbf{math} (76\%), and \textbf{science} (67\%) projects faced difficulties.
The majority of respondents from \textbf{computer science} (56\%) and \textbf{other} (56\%) projects did not face difficulties. 

\begin{table}[!htb]
    \centering
    \caption{Difficulty in determining suitable tolerance (Q22) vs. project domain (Q3)}
    \label{tab:DomainTolerance}
    \begin{tabular}{l|c|c|c|c|c||c}
\thead{} & \thead{CS} & \thead{ENG} & \thead{MA} & \thead{SCI} & \thead{O} & \thead{Total} \\
\hline
Yes & 7 (44\%) &\cellcolor{gray!50} 8 (73\%) &\cellcolor{gray!50}16 (76\%)  &\cellcolor{gray!50}49 (67\%)  &9 (56\%)  &87 (64\%) \\
\hline
No &9 (56\%)  & 3 (27\%) & 5 (24\%) &24 (33\%)  & 7 (44\%)  &50 (36\%) \\
\hline\hline
Total & 16  &11  & 21 & 16 & 73 & 137 \\
\end{tabular} \end{table}

\vspace{12pt}
\noindent
\textbf{RQ5 - Metrics}

\noindent
For test metrics (Table~\ref{tab:TestQualityProjectDomain}), compared with the overall sample, respondents from \textbf{engineering} projects reported a higher frequency of \emph{Branch coverage} (31\% vs. 19\%), a lower frequency of \emph{Statement coverage} (39\% vs. 49\%), and 0\% used \textbf{None}. 
For software quality metrics (Table~\ref{tab:SotwareQualtyProjectDomain}), compared with the overall sample, respondents from \textbf{computer science} projects had a  smaller percentage that selected \emph{None} (27\% vs. 48\%) and a larger percentage that selected \emph{Problems assigned to internal development groups} (23\% vs. 10\%). 
Respondents from \textbf{engineering} projects had a smaller percentage that selected \emph{None} (31\% vs. 48\%) and a larger percentage that selected \emph{Number of defects in current software versions} (38\% vs. 22\%).

\begin{table}[!htb]
    \centering
    \caption{Testing quality metrics (Q24) vs. project domain (Q3)}
    \label{tab:TestQualityProjectDomain}
    \begin{tabular}{l|c|c|c|c|c||c}
\thead{Metrics}  & \thead{CS} & \thead{ENG} & \thead{MA} & \thead{SCI} & \thead{O} & \thead{Total} \\
\hline
Statement & 10 (53\%) & \cellcolor{gray!50}5 (39\%) & 13 (48\%) & 42 (49\%) & 8 (36\%) & 74 (49\%) \\
\hline
Branch & 4 (21\%) & \cellcolor{gray!50}4 (31\%) & 5 (19\%) & 13 (15\%) & 7 (32\%) & 28 (19\%) \\
\hline
Other & 2 (11\%) & 4 (31\%) & 5 (19\%) & 18 (21\%) & 4 (18\%) & 29 (19\%) \\
\hline
None & 3 (16\%) & \cellcolor{gray!50}0 (0\%) & 4 (15\%) & 12 (14\%) & 3 (14\%) & 20 (13\%) \\
\hline\hline
Total & 19 & 13 & 27 & 85 & 22 & 151 \\
\end{tabular} \end{table}

\begin{table}[!htb]
    \centering
    \caption{Software quality metrics (Q25) vs. project domain (Q3)}
    \label{tab:SotwareQualtyProjectDomain}
    \begin{tabular}{p{.2\textwidth}|c|c|c|c|c||c}
\thead{Metrics}  & \thead{CS} & \thead{ENG} & \thead{MA} & \thead{SCI} & \thead{O} & \thead{Total} \\
\hline
Number of defects in current software version & 6 (23\%) & \cellcolor{gray!50}5 (38\%) & 7 (25\%) & 18 (21\%) & 3 (17\%) & 34 (22\%) \\
\hline
Number of defects in previous software versions & 2 (8\%) & 0 (0\%) & 3 (11\%) & 7 (8\%) & 2 (11\%) & 14 (9\%) \\
\hline
Problems assigned to internal development groups & \cellcolor{gray!50}6 (23\%) & 2 (15\%) & 1 (4\%) & 8 (9\%) & 2 (11\%) & 16 (10\%) \\
\hline
Turnaround time on defect corrections & 3 (12\%) & 1 (8\%) & 1 (4\%) & 3 (4\%) & 0 (0\%) & 8 (5\%) \\
\hline
Other & 2 (8\%) & 1 (8\%) & 1 (4\%) & 2 (2\%) & 2 (11\%) & 8 (5\%) \\
\hline
None & \cellcolor{gray!50}7 (27\%) & \cellcolor{gray!50}4 (30\%) & 14 (51\%) & 47 (55\%) & 9 (50\%) & 74 (48\%) \\
\hline\hline
Total & 26 & 13 & 27 & 85 & 18 & 154 \\
\end{tabular} \end{table}

\vspace{4pt}
\noindent
\textbf{RQ6 - Test execution}

\noindent
Regarding test case selection/prioritization (Table~\ref{tab:PrioritizationProjectDomain}), more than half of the respondents from \textbf{computer science} and \textbf{math} projects (56\% and 53\%, respectively) said they use test case selection/prioritization when resources are limited.
Conversely, less than half of the respondents from the \textbf{engineering} and \textbf{science} projects (36\% and 46\%, respectively) did. 

\begin{table}[!htb]
    \centering
    \caption{Test case selection/prioritization (Q30) vs. project domain (Q3)}
    \label{tab:PrioritizationProjectDomain}
    \begin{tabular}{p{.15\textwidth}|c|c|c|c|c||c}
\thead{} & \thead{CS} & \thead{ENG} & \thead{MA} & \thead{SCI} & \thead{O} & \thead{Total} \\
\hline
Yes & \cellcolor{gray!50}9 (56\%) & 4 (36\%) & \cellcolor{gray!50}10 (53\%) & 33 (46\%) & 8 (53\%) & 64 (48\%)\\
\hline
No & 7 (44\%) & \cellcolor{gray!50}7 (64\%) & 9 (47\%) & \cellcolor{gray!50}39 (54\%) & 7 (47\%) & 69 (52\%)\\
\hline\hline
Total & 16 & 11 & 19 & 72 & 15 & 133 \\
\end{tabular} \end{table}

\subsubsection{Q4 - \# of FTEs on the project}
The cross-cutting analysis in this section groups respondents based on their answers to Q4. 
The groups are \textbf{$<$ 1 FTE}, \textbf{1 - 5 FTEs}, \textbf{6 - 20 FTEs}, and \textbf{$>$ 20 FTEs}.

\vspace{4pt}
\noindent
\textbf{RQ1 - Testing characteristics}

\begin{figure}[t]
\centering
\includegraphics[width=0.9\textwidth]{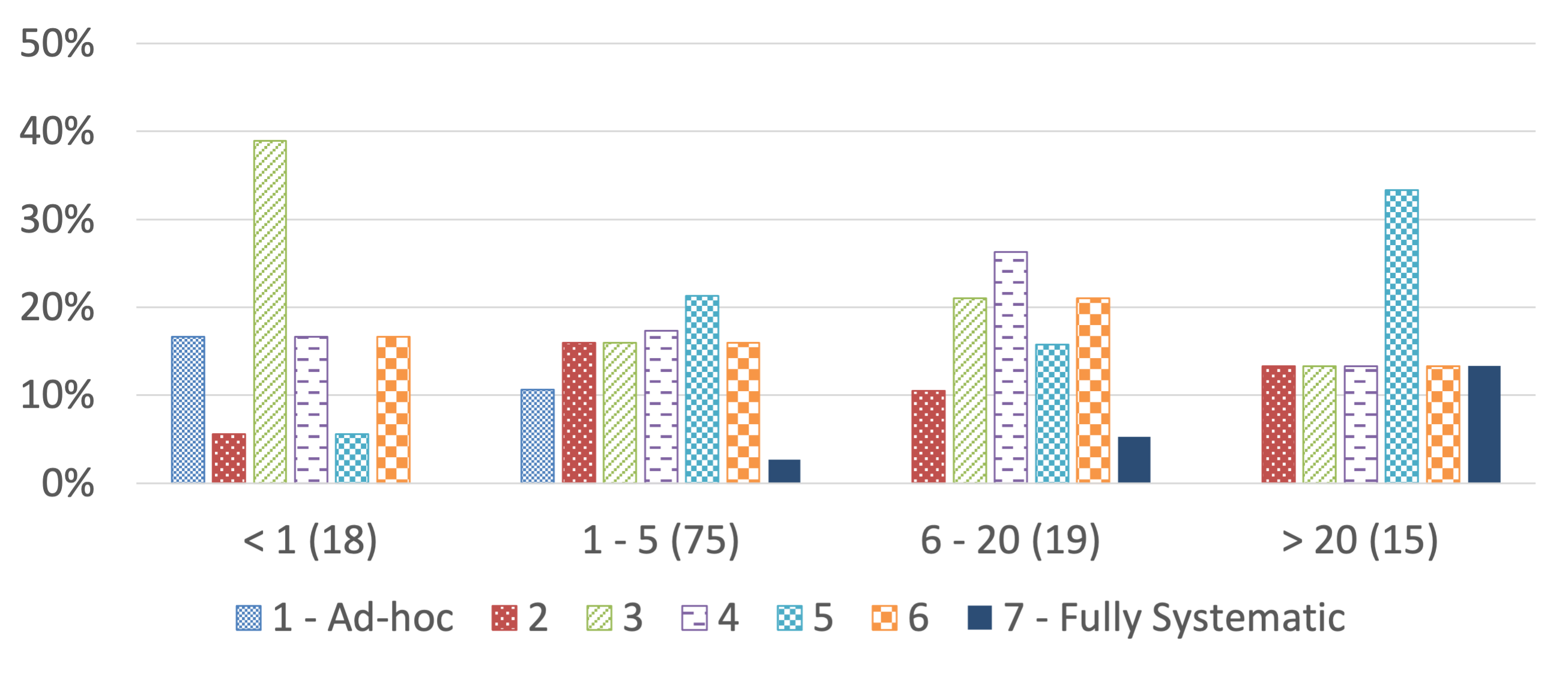}
\caption{Systematic testing process (Q11) vs. \# of FTEs (Q4)}
\label{fig:SystematicVsNoFTEs}
\centering
\end{figure}

\noindent
Respondents from projects with \textbf{$<$ 1 FTE} reported the testing processes were \textit{less systematic} (78\% were on the less systematic end of the scale) and more \textit{manual} (56\% were on the manual end of the scale). See Figure~\ref{fig:SystematicVsNoFTEs} for the data.
Conversely, 60\% of respondents from projects with \textbf{$>$ 20 FTEs} had \textit{documented test requirements} compared with 32\% to 47\% for other projects (Table~\ref{tab:TestReqFTE}).

\begin{table}[!htb]
    \centering
    \caption{Documented test requirements (Q13) vs. \# FTEs on project (Q4)}
    \label{tab:TestReqFTE}
    \begin{tabular}{l|c|c|c|c||c}
\thead{} & \thead{$<$1} & \thead{1-5} & \thead{6-20} & \thead{$>$20} & \thead{Total} \\
\hline
Yes & 6 (32\%) & 29 (38\%) & 9 (47\%) & \cellcolor{gray!50}9 (60\%) & 53 (41\%) \\
\hline
No & 13 (68\%) & 48 (62\%) & 10 (53\%) & 6 (40\%) & 77 (59\%) \\
\hline\hline
Total & 19 & 77 & 19 & 15 & 130 \\
\end{tabular} \end{table}

\vspace{4pt}
\noindent
\textbf{RQ2 - Testing challenges}

\noindent
While \textit{Test Input Design} was the most prevalent challenge for all respondents (Table~\ref{tab:TestChallengesFTEs}), for projects with \textbf{5 or fewer FTEs}, \textit{Evaluating Testing Quality} was the second most common challenge (24\% - 25\%) compared with \textit{Evaluating the correctness of test outputs} for projects with \textbf{6 or more FTEs} (22\% - 25\%).

\begin{table}[!htb]
    \centering
    \caption{Testing challenges (Q15) vs. \# FTEs on project (Q4)}
    \label{tab:TestChallengesFTEs}
    \begin{tabular}{p{2.5cm}|c|c|c|c||c}
\thead{} & \thead{$<$1} & \thead{1-5} & \thead{6-20} & \thead{$>$20} & \thead{Total} \\
\hline
Test Case Design & 17 (30\%) &69 (30\%)  &16 (28\%) &15 (33\%)  &117 (30\%)  \\
\hline
Test Case Execution & 5 (9\%) &20 (9\%)  &2 (3\%) &4 (9\%)  & 31 (8\%) \\
\hline
Evaluating Output Correctness &11 (19\%)  &53 (23\%)  &\cellcolor{gray!50}14 (25\%) &\cellcolor{gray!50}10 (22\%)  &88 (23\%)  \\
\hline
Evaluating Coverage &6 (10\%)  &22 (9\%)  &8 (14\%)  &is 5 (11\%)  &41 (10\%)  \\
\hline
Evaluating Quality &\cellcolor{gray!50}14 (25\%)  &\cellcolor{gray!50}55 (24\%)  &11 (19\%)  &8 (18\%)  &88  (23\%) \\
\hline
Using Test Tools &4 (7\%)  &7 (3\%) &4 (7\%)  & 3 (7\%) &18 (5\%)  \\
\hline
Other & 0 (0\%) &5 (2\%)  &2 (3\%)  &0 (0\%)  &7 (2\%)  \\
\hline\hline
Total & 57  &231  &57  &45  & 390 \\
\end{tabular} \end{table}

\vspace{4pt}
\noindent
\textbf{RQ6 - Test execution}

\noindent
Regarding test case selection/prioritization (Table~\ref{tab:PrioritizationFTEs}), projects with \textbf{$<$ 1 FTE}, \textbf{1 -5 FTEs}, and \textbf{$> 20$ FTEs} had more \textit{No} responses.
Conversely, projects with \textbf{6 - 20 FTEs} had more \textit{Yes} responses.

\begin{table}[!htb]
    \centering
    \caption{Test case selection/prioritization (Q30) vs. \# FTEs on project (Q4)}
    \label{tab:PrioritizationFTEs}
    \begin{tabular}{l|c|c|c|c||c}
\thead{} & \thead{$<$1} & \thead{1-5} & \thead{6-20} & \thead{$>$20} & \thead{Total} \\
\hline
Yes & 5 (31\%) & 41 (49\%) & \cellcolor{gray!50}10 (56\%) & 7 (44\%) & 63 (47\%) \\
\hline
No & \cellcolor{gray!50}11 (69\%) & \cellcolor{gray!50}42 (51\%) & 8 (44\%) & \cellcolor{gray!50}9 (56\%) & 70 (53\%) \\
\hline\hline
Total & 16 & 83 & 18 & 16 & 133 \\
\end{tabular} \end{table}

Regarding how respondents limit test cases when there are challenges in determining the output the most common answer differed (Table~\ref{tab:TestCaseLimtFTEs}): for projects with \textbf{$<$ 1 FTE} and \textbf{1 - 5 FTEs} -- \textit{tests executed less frequently} (56\% and 46\%, respectively), for projects with \textbf{6 to 20 FTEs}, -- \textit{other} (62\%), for projects with \textbf{$>$ 20 FTEs} -- \textit{tests never executed} (50\%).

\begin{table}[!htb]
    \centering
    \caption{How test cases are limited (Q21) vs. \# FTEs on project (Q4)}
    \label{tab:TestCaseLimtFTEs}
    \begin{tabular}{p{3cm}|c|c|c|c||c}
\thead{} & \thead{$<$1} & \thead{1-5} & \thead{6-20} & \thead{$>$20} & \thead{Total} \\
\hline
Those tests are executed less frequently &\cellcolor{gray!50} 5 (56\%) & \cellcolor{gray!50}13 (46\%) & 3 (38\%) & 2 (20\%) & 23 (42\%) \\
\hline
Those tests are never executed & 1 (11\%) & 7 (25\%) & 0 (0\%) & \cellcolor{gray!50}5 (50\%) & 13 (24\%) \\
\hline
Other & 3 (33\%) & 8 (29\%) & \cellcolor{gray!50}5 (62\%) & 3 (30\%) & 19 (34\%) \\
\hline\hline
Total & 9 & 28 & 8 & 10 & 55 \\
\end{tabular} \end{table}

\subsubsection{Q8 - Whether there are any FTEs specifically devoted to testing }
\label{sec:results-cross-Q8}

The cross-cutting analysis in this section groups respondents based on their answers to Q8.
The groups are \textbf{FTEs} and \textbf{No FTEs}.

\vspace{4pt}
\noindent
\textbf{RQ1 - Testing characteristics}

\begin{figure}[!httb]
\centering
\includegraphics[width=0.9\textwidth]{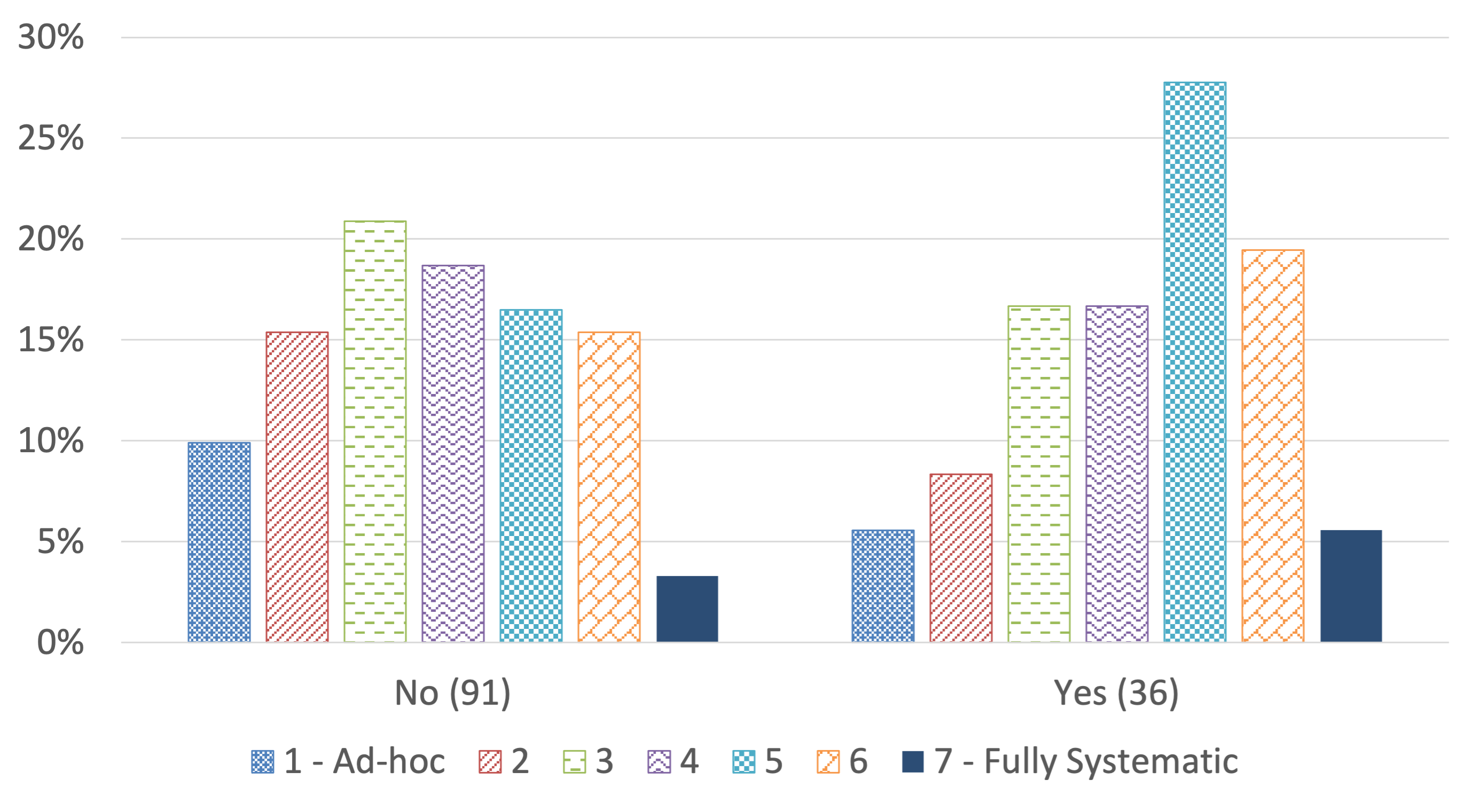}
\caption{Systematic testing process (Q11) Vs. having FTEs devoted to testing (Q8)}
\label{fig:SystematicVsTFTEs}
\centering
\end{figure}

\noindent
Projects in the \textbf{no FTEs} group used a \textit{less systematic}  testing process (65\% were on the less systematic end of the scale) compared with 47\% for the \textbf{FTE} group (Figure~\ref{fig:SystematicVsTFTEs}).
Projects in the \textbf{FTE} group were more likely to have \textit{documented test requirements} (58\% vs. 34\%). See Table~\ref{tab:DocumentedReqVsTFTES} for the data.

\begin{table}[!htb]
    \centering
    \caption{Documented test requirements (Q14) vs. FTEs devoted to testing (Q8)}
    \label{tab:DocumentedReqVsTFTES}
    \begin{tabular}{l|c|c||c}
\thead{} & \thead{No} & \thead{Yes} &\thead{Total} \\
\hline
Yes & 32 (34\%)  & \cellcolor{gray!50}21 (58\%)  & 53 (41\%)  \\
\hline
No & 62 (66\%) &  15 (42\%)  & 77 (59\%)  \\
\hline
Total &94  &36  &130  \\
\end{tabular} \end{table}

\vspace{4pt}
\noindent
\textbf{RQ3 - Test input design}

\noindent
While all respondents listed \textit{Manually} as the most common method of designing test inputs (Table~\ref{tab:InputDeignFTEs}), for those from projects in the \textbf{FTE} group, the second most common answer was \textit{Automatically using tools} (18\%) compared with \textit{Ad hoc} for those from projects in the \textbf{no FTE} group (27\%).

\begin{table}[!htb]
    \centering
    \caption{Test input design (Q16) vs. FTEs devoted to testing (Q8)}
    \label{tab:InputDeignFTEs}
    \begin{tabular}{l|c|c||c}
\thead{} & \thead{No} & \thead{Yes} &\thead{Total} \\
\hline
Manually & 91 (62\%)  & 32 (65\%)  & 123 (63\%)  \\
\hline
Automatically & 13 (9\%)  &\cellcolor{gray!50} 9 (18\%)  & 22 (11\%)  \\
\hline
Ad-hoc &\cellcolor{gray!50} 40 (27\%)  & 7(14\%) & 47 (24\%)   \\
\hline
Other & 2(1\%)  & 1 (2\%) & 3 (1\%)   \\
\hline\hline
Total &146  &49  &195  \\
\end{tabular} \end{table}

\vspace{4pt}
\noindent
\textbf{RQ4 - Test output design}

\noindent
Almost half (47\%) of respondents from projects in the \textbf{FTE} group \emph{often} faced challenges when determining test outputs (Table~\ref{tab:OutputChallengesFTEs}).
The opposite was true for respondents from projects in the \textbf{No FTE} group, 44\% \emph{rarely} faced challenges.
In cases where these challenges limited the number of test cases run (Table~\ref{tab:LimitChallengesFTEs}), 60\% of the respondents from projects in the \textbf{FTEs} group chose \textit{tests executed less frequently} while 40\% of the respondents from projects in the \textbf{no FTE} group chose \textit{Other}.

\begin{table}[!htb]
    \centering
    \caption{Challenges designing test output (Q18) vs. FTEs devoted to testing (Q8)}
    \label{tab:OutputChallengesFTEs}
    \begin{tabular}{l|c|c||c}
\thead{} & \thead{Yes} & \thead{No} & \thead{Total} \\
\hline
Frequently &6 (17\%) & 21 (23\%) & 27 (21\%)  \\
\hline
Often &\cellcolor{gray!50} 17 (47\%)  & 31 (33\%)  & 48 (37\%) \\
\hline
Rarely & 13 (36\%) & \cellcolor{gray!50}41 (44\%)  & 54 (42\%) \\
\hline
Never & 0 (0\%) & 0(0\%)  & 0 (0\%) \\
\hline\hline
Total &36  &93  &129  \\
\end{tabular} \end{table}

\begin{table}[!htb]
    \centering
    \caption{How challenges limit \# tests (Q21) vs. FTEs devoted to testing (Q8)}
    \label{tab:LimitChallengesFTEs}
    \begin{tabular}{l|c|c||c}
\thead{How Limited}  & \thead{Yes} & \thead{No} & \thead{Total} \\\hline
Never Executed & 3(20\%)  & 10 (25\%)  & 13 (24\%)  \\
\hline
Executed Less Frequently & \cellcolor{gray!50}9(60\%)  & 14 (35\%)  &23 (42\%)   \\
\hline
Other & 3(20\%)  & \cellcolor{gray!50}16 (40\%)  & 19 (34\%) \\
\hline\hline
Total &15  &40  &55  \\
\end{tabular} \end{table}

\vspace{4pt}
\noindent
\textbf{RQ6 - Test execution}

\noindent
The most common time for executing test cases is \textit{with a push} for all respondents (Table~\ref{tab:ExecutionTimeFTEs}).
However, for respondents from projects in the \textbf{FTE} group, the second most common time is \textit{Nightly} compared with \textit{at the end of the development cycle} for respondents from the \textbf{no FTE} group.
Respondents from projects in the \textbf{FTE} group were more likely to use test case selection/prioritization when resources are limited than respondents from the \textbf{no FTE} group (Table~\ref{tab:PrioritizationFTEsPresent}).

\begin{table}[!htb]
    \centering
    \caption{When tests executed (Q27) vs. FTEs devoted to testing (Q8)}
    \label{tab:ExecutionTimeFTEs}
    \begin{tabular}{l|c|c||c}
\thead{} & \thead{Yes} & \thead{No} & \thead{Total} \\
\hline
Nightly & \cellcolor{gray!50}13 (23\%) & 14 (8\%) & 27 (12\%)\\
\hline
With a push & 26 (47\%) & 66 (40\%) & \cellcolor{gray!50}92 (41\%) \\
\hline
At the end of development cycle & 11 (20\%) & \cellcolor{gray!50}60 (36\%) & 71 (32\%)\\
\hline
Never & 0 (0\%) & 2 (1\%) & 2 (0.1\%) \\
\hline
Other & 5 (10\%) & 25 (15\%) & 30 (14\%)\\
\hline\hline
Total & 55 & 167 & 222 \\
\end{tabular} \end{table}

\begin{table}[!htb]
    \centering
    \caption{Use of test case selection/prioritization (Q30) vs. FTEs devoted to testing (Q8)}
    \label{tab:PrioritizationFTEsPresent}
    \begin{tabular}{l|c|c||c}
\thead{} & \thead{Yes} & \thead{No} & \thead{Total} \\
\hline
Yes & \cellcolor{gray!50}23 (37\%) & 12 (17\%) & 35 (26\%) \\
\hline
No & 40 (63\%) & \cellcolor{gray!50}58 (83\%) & 98 (74\%) \\
\hline\hline
Total & 63 & 70 & 133 \\
\end{tabular} \end{table}

\subsubsection{Q10 - Percentage of development time spent testing}
\label{sec:results-cross-Q10}

The cross-cutting analysis in this section groups respondents based on their answers to Q10.
The groups are \textbf{0\%}, \textbf{1-25\%}, \textbf{26-50\%}, \textbf{50-75\%}, \textbf{76-100\%}, and \textbf{Unknown}. 

\vspace{4pt}
\noindent
\textbf{RQ1 - Testing characteristics}

\begin{figure}[t]
\centering
\includegraphics[width=0.9\textwidth]{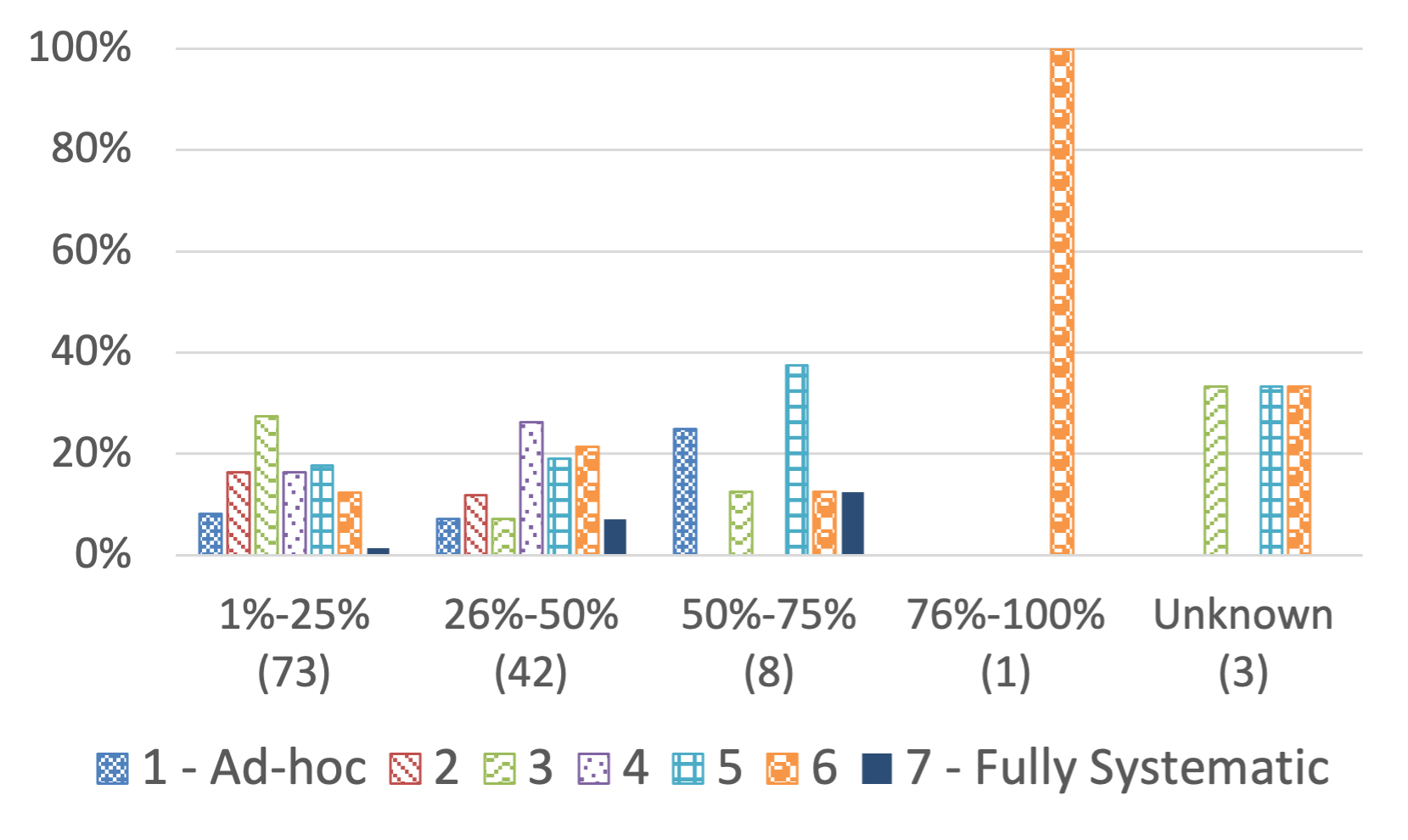}
\caption{Systematic testing process (Q11) Vs. percentage of time spent on testing (Q10)}
\label{fig:TestTimeVsSystematic}
\centering
\end{figure}

\noindent
Projects that spend less time on testing activities also tend to follow a \textit{less systematic} process overall.
Specifically, 68\% of respondents from projects in the \textbf{1-25\%} group were on the \textit{less systematic} end of the scale, compared with 52\% from the \textbf{26-50\%} group, and 37\% from the \textbf{50-75\%} group (Figure~\ref{fig:TestTimeVsSystematic}).
Projects that spend less time on testing also tend to have \textit{fewer documented testing requirements}. 
Specifically, 29\% of respondents from projects in the \textbf{1-25\%} group had \textit{documented test requirements}, compared with 52\% from the \textbf{26-50\%} group, and 66\% from the \textbf{50-75\%} group (Table~\ref{tab:DocumentedReqVsTestTime}).

\begin{table}[!htb]
    \centering
    \caption{Documented test requirements (Q14) vs. percentage of time spent on testing (Q10)}
    \label{tab:DocumentedReqVsTestTime}
    \begin{tabular}{p{.15\textwidth}|c|c|c|c|c||c}
\textbf{Metrics} & \rot{\textbf{1\%-25\%}} & \rot{\textbf{26\%-50\%}} & \rot{\textbf{51\%-75\%}} & \rot{\textbf{75\%-100\%}} & \rot{\textbf{Unknown}} & \rot{\textbf{Overall}} \\
\hline
Yes & \cellcolor{gray!50}22 (29\%) & 22 (52\%) & 6 (67\%) & 1 (100\%) & 2 (67\%) & 53 (41\%) \\
\hline
No & 53 (71\%) & 20 (48\%) & 3 (33\%) & 0 (0\%) & 1 (33\%) & 77 (59\%) \\
\hline\hline
Total & 75 & 32 & 9 & 1 & 3 & 133 \\
\end{tabular} \end{table}

\vspace{4pt}
\noindent
\textbf{RQ5 - Metrics}
Regarding the percent of time spent testing (Table~\ref{tab:TestQualityTestTime}) all respondents from projects in the \textbf{50-75\%} and \textbf{76-100\%} groups some type of \textit{test quality metrics}. 
In addition, the respondents from projects in the \textbf{51-75\%} group tended to use \textit{Branch} coverage more than the overall average (33\% vs. 19\%) and \textit{Statement coverage} less than the overall average (42\% vs. 49\%).

\begin{table}[!htb]
    \centering
    \caption{Testing quality metrics (Q24) vs. percent of time spent testing (Q10)}
    \label{tab:TestQualityTestTime}
    \begin{tabular}{p{.15\textwidth}|c|c|c|c|c||c}
\textbf{Metrics} & \rot{\textbf{1\%-25\%}} & \rot{\textbf{26\%-50\%}} & \rot{\textbf{51\%-75\%}} & \rot{\textbf{75\%-100\%}} & \rot{\textbf{Unknown}} & \rot{\textbf{Overall}} \\
\hline
Statement & 41 (47\%) & 24 (50\%) & \cellcolor{gray!50}5 (42\%) & 1 (100\%) & 3 (75\%) & 74 (49\%) \\
\hline
Branch & 15 (17\%) & 8 (17\%) & \cellcolor{gray!50}4 (33\%) & 0 (0\%) & 1 (25\%) & 28 (19\%) \\
\hline
Other & 19 (22\%) & 7 (15\%) & 3 (25\%) & 0 (0\%) & 0 (0\%) & 29 (19\%) \\
\hline
None & 11 (13\%) & 9 (19\%) & \cellcolor{gray!50}0 (0\%) & \cellcolor{gray!50}0 (0\%) & 0 (0\%) & 20 (13\%) \\
\hline\hline
Total & 86 & 48 & 12 & 1 & 4 & 151 \\
\end{tabular} \end{table}

\vspace{4pt}
\noindent
\textbf{RQ6 - Test execution}
As the percentage of time spent on testing increases over 50\%, so does the likelihood that respondents use some type of \textit{test case selection or prioritization} (48\% for those in the \textbf{1 - 25\%} group, 38\% for those in the \textbf{26 - 50\%} group, 75\% for those in the \textbf{50\% - 75\%} group, and 67\% for those in the \textbf{76 - 100\%} group) (Table~\ref{tab:TestPrioritizationTestTime}).

\begin{table}[!htb]
    \centering
    \caption{Use of test case selection/prioritization (Q30) vs. percent of time spent testing (Q10)}
    \label{tab:TestPrioritizationTestTime}
    \begin{tabular}{p{.15\textwidth}|c|c|c|c|c||c}
\textbf{Metrics} & \rot{\textbf{1\%-25\%}} & \rot{\textbf{26\%-50\%}} & \rot{\textbf{51\%-75\%}} & \rot{\textbf{75\%-100\%}} & \rot{\textbf{Unknown}} & \rot{\textbf{Overall}} \\
\hline
Yes & 38 (48\%) & 16 (38\%) & \cellcolor{gray!50}6 (75\%) & \cellcolor{gray!50}1 (50\%) & \cellcolor{gray!50}2 (67\%) & 63 (47\%) \\
\hline
No & 41 (52\%) & 26 (62\%) & 2 (25\%) & 1 (50\%) & 1 (33\%) & 71 (53\%) \\
\hline\hline
Total & 79 & 42 & 8 & 2 & 3 & 133 \\
\end{tabular} \end{table}

\section{Discussion}
\label{sec:discussion}
In this section, we answer each research question based on the results described in Section~\ref{sec:results}.

\subsection{RQ1: Characteristics of the testing process}
None of the previous studies we discussed in Section~\ref{sec:researchQuestions}~\citep{eisty2022,10.1007/978-3-030-50436-6_33,KANEWALA20141219, 8588655, HEATON2015207} looked at the level of automation and the ad-hoc/systematic nature of the research software testing process. 
The research software testing process varies widely from ad-hoc to systematic. 
Similarly, the level of automation varies widely with a slight skew toward more automation. 
However, research software developers often equate test automation to the automatic execution of tests. 
This misunderstanding may be the reason for the slight skew toward more automation.

The most common type of testing used in research software is unit testing, closely followed by regression testing. 
Some projects also use integrating testing, system testing, and acceptance testing. 
While these findings are similar to the findings of our earlier survey~\citep{eisty2022},
they differ from the findings of a survey of industrial developers, which found most of the budget allocated to system tests and less allocated for unit tests~\citep{ISTQB2018}. 
Another industry survey shows an increased use of acceptance testing within the past several years~\citep{surveyofpractice}. 
While the testing types used by the research software developers did not change over the several years between our previous survey and this current survey, the emphasis on unit testing is significantly higher among research software developers than among industry developers. 
This result may be due to a lack of knowledge about the importance of other testing techniques, such as acceptance testing and system testing, or the practical difficulties of applying other types of testing techniques, such as the oracle problem in system testing or budgetary constraints for acceptance testing. 

In addition, most research software projects do not have documented test requirements. 
Employing this practice could help research software projects because it is useful for improving the overall quality of the testing process.

\subsection{RQ2: Challenges faced when testing research software}
\textit{Test case design} is the most challenging testing task in research software projects.
This result is consistent with previous studies~\citep{KANEWALA20141219,eisty2022}. 
However, none of the previous studies specifically examined why these testing tasks are challenging.
Our survey shows test case design was challenging due to difficulties in designing good test cases, oracle problems, and the lack of resources for testing. 
Addressing these challenges could significantly enhance the research software development process by helping developers overcome the biggest challenges to thoroughly testing their software and producing high-quality results. 

A survey of industry practitioners identified test automation, documentation, and collaboration as the most common challenges~\citep{ISTQB2018,world_quality_report_2021}. 
None of these challenges were provided as multiple-choice options in our survey.
We provided more granular testing tasks as options to help identify specific testing challenges that create bottlenecks in the testing process of research software. 
Interestingly, the research software developers did not mention the above three challenges in the ``Other'' challenge option. 
Thus, our results show that research software developers face significant challenges in more mainstream testing tasks than developers in industry.

\subsection{RQ3: Test input design}
None of the previous studies looked in-depth into what specific methods are used for test input design is done in research software development. 
Respondents most commonly used a \emph{manual} approach for test input design.
Respondents did not mention existing techniques focused on input generation, such as combinatorial testing~\citep{10.1145/1883612.1883618} and fuzz testing~\citep{ManesFuzz2021}. 
On the other hand, a survey done on industry developers reported using various techniques, including pair-wise testing (a form of combinatorial testing), equivalence partitioning, boundary value analysis, and coverage-based testing~\citep{ISTQB2018,PractintionersViews}. 
Another survey done on industry developers reported that many use automated test data generation, automated test script generation, and model-based test generation for automated test input design~\citep{world_quality_report_2021}.
Research software developers may be unaware of these techniques, and/or they might not be effective for research software because of their complexity. 
There is a need to evaluate the effectiveness of these techniques for research software.

\subsection{RQ4: Determining the expected output}
None of the previous studies looked in-depth into
challenges associated with determining the expected output in research software development. 
The majority of challenges in determining outputs resulted from the oracle problem. 
The underlying causes mentioned by the respondents for the oracle problem included code complexity,  non-determinism, and difficulties in evaluating test outputs. 
Most respondents faced these challenges either \emph{often} or \emph{frequently}. 
Therefore, addressing these challenges would help to improve the effectiveness of the scientific software testing process. 
In addition, due to these challenges, developers either never executed those tests or executed them less frequently, which can directly impact the quality of research software. 
Techniques such as metamorphic testing~\citep{ChenMetamorphic2019} effectively handle the oracle problem. 
However, none of the respondents mentioned it.
This omission highlights a gap in adapting such techniques for the research software domain. 
There are many possible reasons for this gap, including a lack of knowledge about these techniques among the research software developers, a lack of tool support to ease the adoption of these techniques in the research software development process, or a lack of resources to adopt techniques. 
Thus, an important future step is to understand the underlying cause for the lack of adoption of these techniques. 

\subsection{RQ5: Metrics for software quality and test quality}
More respondents used metrics to measure test quality than used metrics to measure overall software quality.
The most common measure of quality is \textit{Statement Coverage}. 
For overall quality, a plurality of respondents used \textit{number of defects in the current software version}. 
Industry surveys report using similar quality metrics on smart/automated dashboards to enable continuous quality monitoring~\citep{world_quality_report_2021}.

\subsection{RQ6: Test Execution}
As expected, most respondents executed tests either with a push or at the end of the development cycle. 
Only one respondent \textit{never} executed tests. Interestingly, no respondents mentioned test-first development. 

As expected, most respondents selected 
\emph{no} for using test case selection and prioritization. 
Even when respondents use them, they typically use a \emph{Manual} approach. 
In contrast, an industry-focused survey reported using artificial intelligence (AI) techniques to prioritize test cases~\citep{world_quality_report_2021}. 
While utilizing automated techniques for test selection/prioritization could be beneficial, especially considering the lack of resources for testing research software, incorporating these techniques into the development process would require overhead, including providing developers with the necessary knowledge and gaining access to necessary tools.

\subsection{RQ7: Testing Tools Used}
\label{sec:discussion:RQ7}
When we asked respondents to list the tools they used for input generation, the ones they listed were not necessarily used for input design but rather general testing frameworks and continuous integration platforms. 
This misunderstanding shows their lack of knowledge of the specific tools used for input generation. 
While input generation tools exist, e.g. Pynguin~\citep{LukasczykPynguin2022}, Pex~\citep{TillmannPex2008} and TackleTest~\citep{TackleTestTzoref2022}, the respondents did not mention any of them.

Relative to tools used to measure quality, no consistent pattern emerged. 
The most popular answer given was \textit{Codecov}. 
However, it represented less than 20\% of the responses.

More than 80\% of the respondents used a tool to execute their tests.
Of those, there was a relatively even split between those who use \textit{GitHub Actions} and those who use \textit{Pytest}.
However, when asked about test case selection or prioritization, just under half of the respondents did not use any tools (e.g. they performed this task manually).

When asked about other tools used for testing, the responses included many tools, each of which was only mentioned by a few people. 
While this result may be due to the diversity of testing needs in research software, this situation makes it difficult for research software developers to share knowledge and hinders the sustainable use of tools.

Considerable differences can be seen in the diversity of tool usage of the industry developers. 
They report using tools for various testing tasks such as defect tracking, test automation, test execution, test design, and unit testing~\citep{ISTQB2018}.
They also commonly use tools at all levels of testing~\citep{surveyofpractice}. 
Another industry survey reported most industry developers find \emph{the tools and methods required for test activities are sufficient and available}~\citep{world_quality_report_2021}. 
Due to cost constraints, these testing tools may be unavailable to the research software developers, or the lack of knowledge about these tools may hinder their adoption in the research software domain.

\subsection{RQ8: Desired Features of Testing Tools}
One potential reason for the lack of a standardized tool set, as discussed in Section~\ref{sec:discussion:RQ7}, is the limitations of those tools.
Most commonly, the respondents reported the tools were \textit{missing important features}.
Similarly, the most required feature respondents wanted in testing tools was \textit{ease of use}.
Together these results suggest an improved toolchain with appropriate functionality and ease of use may help consolidate the community to a smaller number of tools.

\subsection{RQ9: Influence of Demographics on Results}
Surprisingly, the testing process for respondents from \textbf{computer science} projects was skewed towards manual testing. 
This result could result from a lack of knowledge by respondents outside of computer science who consider test execution equivalent to the overall testing process.
In other words, if they use an automated execution environment, the respondents may have said their testing process was automated.

Regardless of the domain \emph{Test input design}, \emph{Evaluating test quality}, and \emph{Evaluating the correctness of the test outputs} were among the top three challenges, reported by 71\% to 82\% of respondents. 
Therefore, testing tools or techniques developed for research software should address these challenges. 
In particular, because \emph{evaluating the correctness of the test outputs} is not commonly faced by traditional software, existing tools do not sufficiently alleviate this challenge. 

As expected, respondents from projects in the \textbf{engineering} and \textbf{science} domains commonly faced the challenge of \emph{determining the expected output} of a test case. 
Particularly for respondents from projects in the \textbf{science} domain, this challenge led to \textit{limiting the number of test cases executed}, which can reduce the efficiency of fault detection. 
Therefore, these developers can benefit from adopting testing techniques such as metamorphic testing~\citep{kanewalaMetamorphicCiSE2019} that can be used to conduct testing when the output is unknown. In addition, respondents from the \textbf{Engineering} and \textbf{Computer Science} domains seem better at using metrics than the respondents from other domains. This is an area where training can help increase the use of metrics in other domains.

The number of FTEs in a project impacted the testing practices used. 
Smaller teams used less systematic and more manual testing processes. 
The same observations could be made regarding the testing process of teams with dedicated FTEs for testing. 
Therefore, increasing the human resources for developing and testing research software projects will likely affect the testing processes used. 
In addition, the presence of dedicated FTEs for testing showed a clear difference in the testing practices used, such as \textit{using automated tools for test input design}, \textit{using test metrics}, and \textit{using test selection/prioritization techniques}. 
Thus the presence of dedicated FTEs should help projects use more tools and techniques and improve testing effectiveness.    

Overall, the results from Section~\ref{sec:results-cross-Q8} suggest the amount of time spent on testing is a good proxy for whether projects follow good software engineering practices. 
The lack of dedicated FTEs for testing may lead to limited test sets. 
The results in Section~\ref{sec:results-cross-Q10} also illustrate instances where respondents on projects with more time devoted to testing tend to be more likely to follow standard software engineering processes.

\section{Practical Implications}
\label{sec:implications}
The practical implications of this study impact multiple stakeholders involved in research software development, including research software developers, users, project managers, funding agencies, and tool developers. 
By acting on these implications, stakeholders can contribute to the development of higher-quality, more reliable research software.
The implications are::

\begin{itemize}
    \item \textbf{Compromised software quality}: the study highlights several instances where the testing process is negatively impacted by various testing challenges, lack of resources, and lack of knowledge. This could lead to compromising the research software's quality, negatively impacting the users and other stakeholders.
    \item \textbf{Improving Testing Practices:} The study highlights the need for improving testing practices in research software projects. Research software developers and project managers can use the insights to implement more systematic testing processes by documenting test requirements, using automated testing techniques, and prioritizing test case design. As a result of improved testing practices, the users of research software will benefit from more correct software.
    \item \textbf{Addressing Testing Challenges:} By identifying common challenges in testing research software, such as test case design and determining expected outputs, stakeholders can focus on addressing these specific issues. These solutions involve providing training and resources on effective test case design strategies and developing tools tailored to address the oracle problem in research software.
    \item \textbf{Enhancing Tool Support:} Tool developers can leverage these findings to develop testing tools that meet the specific needs of research software developers. These tools need to prioritize features, including ease of use, support for test input design, and addressing the oracle problem.
    \item \textbf{Resource Allocation:} Project managers and funding agencies can use the results to better allocate resources for testing activities in research software projects. In addition, an understanding of the impact of team size and dedicated testing resources on testing practices should inform staffing and budget allocation decisions.
    \item \textbf{Community Building and Knowledge Sharing:} The study underscores the importance of knowledge sharing and community building within the research software development community. Initiatives aimed at sharing best practices, lessons learned, and tools for testing research software can foster collaboration and accelerate improvements in testing practices.
    \item \textbf{Educational Initiatives:} Academic institutions and training programs can integrate insights from this study into their curriculum to better prepare students and researchers for testing research software. By emphasizing the importance of testing and providing training on effective testing techniques, educational initiatives can contribute to the overall quality and reliability of research software.
\end{itemize}

\section{Threats to Validity}
\label{sec:threatsToValidity}
This section describes the threats to validity of the study.

\subsection{Internal Threats}
The primary threat to internal validity relates to the software engineering terminology used in the survey.
If respondents, who were typically not traditional software engineers, had a different understanding or different definitions of the terms, then our conclusions may be less reliable.
To reduce this threat, we conducted interviews and pilot studies to validate the survey (Section~\ref{sec:SurveyDesign}).

\subsection{External Threats}
If the survey respondents are not representative of the population of research software developers, the results are less generalizable.
In addition, because we cannot measure the size of a community as diverse as research software developers, it is possible that our sample is not representative.
To reduce this threat, we used multiple recruitment approaches, targeting large groups of research software developers.
While it is clear that most respondents self-identified as RSEs (82\%), who clearly are research software developers, the fact they took time to respond to a survey about testing and provided detailed responses suggests the results could be biased towards people who are already predisposed towards the use of testing.

\subsection{Construct Threats}
The primary construct validity threat is the respondents may have misunderstood the questions. 
To reduce this threat, we took great care in writing the survey questions and conducted interviews and pilot studies to validate the survey (Section~\ref{sec:SurveyDesign})

\subsection{Conclusion Threats}
It is possible that, with additional information, other conclusions could be drawn from the data gathered in this study.
We rely on the respondents' perceptions of software testing, which may not match reality.

\section{Conclusions}
\label{sec:conclusions}
In this study, we explore research software developers' perspectives on software testing. 
Industrial testing tools are often inadequate for use in research software due to their complexity, domain-specific nature, and continuous evolution. 
Research software involves specialized algorithms and frequent updates and focuses more on validation than verification, requiring flexible and custom testing approaches. 
The lack of standardization and non-deterministic outputs further complicate traditional testing methods, making specialized strategies essential.

Based on our research questions, we designed a survey. 
We utilized various dissemination channels focused on research software developers to ensure the survey reached a broad and diverse sample.
We received 131 valid responses, which we analyzed using quantitative and qualitative methods.

Our results show the level of automation and the systematic nature of the testing process for research software projects vary widely. 
Certain demographic characteristics such as the domain, number of FTEs, and the availability of dedicated FTEs for testing also affect the testing process used. 
Allocation of human resources to the testing and to the project, in general, helps to increase the effectiveness of the testing process.

Nevertheless, regardless of the demographics, the respondents identified test case design as the biggest testing challenge. 
Therefore, providing training on tools and techniques for effective input design may be valuable to research software developers. 

In addition, research should investigate how existing testing techniques can be used to perform testing in the presence of the oracle problem.
Because this problem is common in many scientific domains, it has resulted in the execution of fewer test cases.

Respondents reported a large number of tools used for various testing tasks. 
However, respondents reported using tools for tasks other than those for which they were originally designed. 
These tools also may not address specific testing challenges faced by research software projects. 
This result indicates a need for testing tools that address specific needs of research software testing.
The research software development community could then adopt such tools, which would increase testing knowledge through continued and sustained use of appropriate testing tools instead of one-off and ad-hoc usage of tools.   

Key conclusions from the study reveal several critical insights into research software testing:

\begin{itemize}
    \item \textbf{Testing Practices Lag Behind Industry:} Research software testing is less systematic and less automated than in industry, with significant gaps in knowledge and tool adoption.
    \item \textbf{Unique Challenges:} The oracle problem and test case design are major challenges, especially in complex, non-deterministic systems.
    \item \textbf{Resource Limitations:} Smaller teams and limited budgets lead to ad-hoc testing, compromising software quality and reliability.
    \item \textbf{Need for Specialized Tools:} Current tools do not fully address the unique needs of research software, highlighting a demand for more tailored solutions.
    \item \textbf{Cultural Shift Needed:} Testing is often an afterthought in research software. A change in mindset is required, integrating testing as a core part of development.
    \item \textbf{Educational Opportunities:} Integrating software testing education into research curricula can help close the knowledge gap and improve future practices.
    \item \textbf{Community Building:} Strengthening collaboration and knowledge sharing in the research software community can help address tool fragmentation and improve sustainability.
\end{itemize}

To conclude, our study highlights the unique challenges and gaps in testing for research software. 
Research software's complexity and evolving nature necessitate specialized, flexible testing approaches. 
Despite the varied levels of automation and systematic testing available, the common hurdles, such as the oracle problem and challenges in test case design, persist across demographics. 
Addressing these issues requires tailored tools, increased resource allocation, and a cultural shift toward integrating testing into the development process. 
By fostering education and community collaboration, we can enhance the effectiveness and sustainability of research software testing practices.

\section*{Data Availability Statements}
\label{sec:dataAvailability}
Due to IRB restrictions, the data generated during the current study are available from the corresponding author on reasonable request. 
\section*{Compliance With Ethical Standards}
\label{sec:compliance}
\textit{Conflict of Interest:} None \\

\noindent
\textit{Funding:} None \\

\noindent
\textit{Ethical approval:} This study received IRB approval from the University of North Florida (IRB\#1660344-1). \\

\noindent
\textit{Informed consent:} Participants had to complete an informed consent form before beginning the survey. \\

\noindent
\textit{Author Contributions:}
All authors contributed to the conception and design of the study.
All authors participated in the data analysis process.
All authors contributed to the manuscript. 
All authors read and approved the final manuscript. 
\bibliographystyle{spbasic}
\bibliography{refs}

\appendix
\section{Appendix - Question Mapping}
\label{appendix_a}
Table~\ref{tab_question_mapping} provides a mapping between the survey questions and the demographic questions to indicate which are included in the paper and which are omitted because the demographics had no impact on the result.
For the ones included in the paper, we place a \checkmark. 
For the ones excluded due to lack of demographic impact, we place a `\textbf{O}'.

\begin{table}[!htb]
    \centering
    \begin{tabular}{c||c|c|c|c|}
\thead{} & \thead{Q3} & \thead {Q4} & \thead {Q8} & \thead{Q10} \\
\hline
Q11 & \textbf{X} & \checkmark & \checkmark & \checkmark \\
\hline
Q12 & \checkmark & \textbf{O} & \textbf{O} & \textbf{O} \\
\hline
Q13 & \textbf{O} & \textbf{O} & \textbf{O} & \textbf{O} \\
\hline
Q14 & \textbf{O} & \checkmark & \checkmark & \checkmark \\
\hline
Q15 & \textbf{O} & \checkmark & \textbf{O} & \textbf{O} \\
\hline
Q16 & \textbf{O} & \checkmark & \checkmark & \textbf{O} \\
\hline
Q18 & \checkmark & \textbf{O} & \checkmark & \checkmark \\
\hline
Q20 & \checkmark & \textbf{O} & \textbf{O} & \textbf{O} \\
\hline
Q21 & \checkmark & \checkmark & \checkmark & \textbf{O} \\
\hline
Q22 & \checkmark & \textbf{O} & \textbf{O} & \textbf{O} \\
\hline
Q24 & \checkmark & \textbf{O} & \textbf{O} & \checkmark \\
\hline
Q25 & \checkmark & \textbf{O} & \textbf{O} & \textbf{O} \\
\hline
Q27 & \textbf{O} & \textbf{O} & \checkmark & \textbf{O} \\
\hline
Q28 & \textbf{O} & \textbf{O} & \checkmark & \checkmark \\
\hline
Q30 & \checkmark & \checkmark & \textbf{O} & \checkmark \\
\hline
\end{tabular}     \caption{Mapping of Questions to Demographics}
    \label{tab_question_mapping}
\end{table} 
\end{document}